\documentclass[10pt]{article}   
\pdfoutput=1
\usepackage{jheppub}
\usepackage{amsmath,amssymb,amsfonts,mathbbol,graphicx,slashed,color,amsthm, mathtools, upgreek, enumerate, tensor}

\usepackage{subcaption}
\usepackage{setspace}
\usepackage[export]{adjustbox}
\usepackage{arydshln}
\usepackage[dvipsnames]{xcolor}
\usepackage{physics}
\usepackage{hyperref}
\usepackage{comment}
\graphicspath{{pics/}}
\usepackage{breqn}

\usepackage{comment}

\usepackage{algorithm}
\usepackage{algpseudocode}

\allowdisplaybreaks

\colorlet{darkblue}{blue!70!black}

\colorlet{darkgreen}{green!50!black}
\colorlet{darkred}{red!50!black}

\newcommand{\mc}{\mathcal}

\def\bea{\begin{eqnarray}}
\def\eea{\end{eqnarray}}
\def\be{\begin{equation}}
\def\ee{\end{equation}}

\title{Bulk-local $dS_3$ holography:  the Matter with $T\bar T+\Lambda_2$
}

\author[a]{Gauri Batra,}
\author[a]{G. Bruno De Luca,}
\author[a]{Eva Silverstein,}
\author[b]{Gonzalo Torroba,}
\author[a]{Sungyeon Yang}
\affiliation[a]{Stanford Institute for Theoretical Physics, 382 Via Pueblo, Stanford, CA 94305}
\affiliation[b]{Centro At\'omico Bariloche and CONICET, Bariloche, RN, Argentina}
 
 \vspace{5mm}

\vspace{1cm}

\abstract{

We propose an algorithm which builds a concrete dual for large-radius 3d de Sitter with a timelike York boundary for both gravity and bulk effective fields.  This generalizes the solvable $T\bar T+\Lambda_2$ deformation,  whose finite real spectrum accounts for the refined Gibbons-Hawking entropy as a microstate count while reproducing the radial static patch geometry. The required generalization to produce approximately local boundary conditions for bulk quantum fields requires a scheme for defining double-trace operators dual to deformed boundary conditions to realize the finite timelike boundary, valid at finite N.   By starting with a small stint of a pure $T\bar T$ trajectory, the theory becomes finite, enabling well-defined subtractions to define the double-trace deformation so as to match the large-N prescription of Hartman, Kruthoff, Shaghoulian, and Tajdini to good approximation.  We incorporate the matter effecting an uplift from negative to positive cosmological constant, and analyze the effect of matter on the energy spectrum of the theory arising from time-dependent bulk excitations.  This validates the cosmic horizon $dS_3$ microstate count for large-radius $dS_3$ holography, embedding $T\bar T+\Lambda_2$ concretely into a larger theory consistent with bulk locality for matter fields.  We comment briefly on potential upgrades to four dimensions and other future directions.          

 }

\begin{document}

\maketitle
\parskip=10pt

\section{Introduction and high-level summary}\label{sec:intro}

Recently the de Sitter entropy in three dimensional gravity \cite{Gibbons:1977mu, Anninos:2020hfj},\footnote{The logarithmic term  was calculated in \cite{Anninos:2020hfj} for unbounded de Sitter, agreeing with the result in \cite{Sen:2012dw} for the corresponding AdS black hole.  Our case with a timelike boundary requires the analogous calculation in the presence of the boundary; the boundary conditions can change the factor in front of the log (as discussed for the case of an observer in \cite{Chandrasekaran:2022cip}).  As we review later in this paper, for the cosmic horizon patch the matching between the AdS and dS parts of the deformation trajectory in our construction involves the boundary skirting the horizon, a region insensitive to the bulk cosmological constant \cite{Coleman:2021nor, Silverstein:2022dfj}.  This ensures that the AdS black hole and dS results agree including the refinements to the Gibbons-Hawking entropy of the sort computed in \cite{Anninos:2020hfj}.} 
\be
S\simeq {\cal A}/4 G_N - c_L \log({\cal A}/4 G_N)
\ee  
along with the radial geometry of the static patch has been captured by an explicit boundary dual defined by the solvable $T\bar T+\Lambda_2 $ deformation \cite{Coleman:2021nor, Gorbenko:2018oov, Lewkowycz:2019xse} of a holographic seed CFT.  In this formulation, there is a clear explanation for the finiteness of the de Sitter entropy:  it is a direct consequence of the finite real spectrum of the $T\bar T$-type theory \cite{Smirnov:2016lqw, Dubovsky:2018bmo, Cavaglia:2016oda}.  This construction includes a timelike non-asymptotic non-gravitational boundary on the gravity side, in whose presence the thermodynamics matches that of ordinary quantum systems \cite{Miyashita:2021iru, Draper:2022ofa, Banihashemi:2022htw, Banihashemi:2022jys}.\footnote{Various subtleties with arbitrary Dirichlet boundary conditions arise in $4d$ General Relativity (see e.g. \cite{Marolf:2012dr, Anderson:2006lqb, Andrade:2015gja, An:2021fcq, Witten:2018lgb, Marolf:2022jra, Marolf:2022ntb}).  
In this work we will stick to 3 external dimensions in the special case of a cylinder boundary within which we solve the matter plus gravity field equations to ensure a good initial boundary value problem.  In Lorentzian signature in 3 space-time dimensions, there is no pileup of states with large extrinsic curvature (cf \cite{Witten:2018lgb, Iliesiu:2020zld, Stanford:2020qhm}), as can be seen by  the finiteness of the real spectrum related to the square root appearing in the dressed Hamiltonian (derived explicitly for Dirichlet boundaries in \cite{Kraus:2022mnu, Ebert:2022cle}). We comment on this in \S\ref{sec-grav-sector} and on potential generalizations to 4d with conformal boundary conditions \cite{Anderson:2006lqb, Witten:2018lgb, Anninos:2023epi, Anninos:2024wpy, Liu:2024ymn} at the end.}     The existence of such boundaries in quantum gravity with matter is a priori an open question, whose resolution may involve direct gravity-side analysis in string theory (e.g. along the lines of \cite{Ahmadain:2022tew, Silverstein:2022dfj})\footnote{We thank A. Ahmadain, R. Khan, and A. Wall for discussions of this point}, or may be established indirectly by formulating a boundary dual matching the low energy bulk theory in a bounded spacetime.\footnote{See e.g. \cite{Manschot:2022lib} for a recent analysis of a $T\bar{T}$ deformation of the Maldacena-Strominger-Witten CFT.}  

The boundary theory \cite{Coleman:2021nor} independently computes universal features such as the horizon entropy and geometry, agreeing with that predicted by bulk semiclassical gravity.  But it does not accurately capture important model-dependent details. These include the  fine structure of the band of energies corresponding to horizon entropy, the details of the uplift from AdS to dS spacetime, and local bulk matter excitations contained within the bounded patch.  These details are subleading in the calculation of the total state count (entropy) and the large-radius geometry, but are important to obtain in a complete theory of de Sitter quantum gravity.  
It is the goal of this paper to fill in this gap and present a concrete proposal for constructing a boundary theory formulating the large-radius 3d de Sitter static patch with timelike boundary.\footnote{This accords with recent work exhibiting simplifications of bounded patch thermodynamics in terms of the quasilocal energy \cite{Banihashemi:2022htw, Banihashemi:2022jys} and with many works over the years such as \cite{Anninos:2011af, Anninos:2011jp, Anninos:2022ujl, Alishahiha:2004md, Dong:2010pm,  Anninos:2023epi, Shaghoulian:2022fop, susskind:2022bia,  Narovlansky:2023lfz, Rahman:2024vyg} analyzing static patch and dS/dS patch holography, including the role of timelike boundaries or observers and the maximal mixing of the correspondinig microstates in global de Sitter  \cite{Banks:2000fe,Bousso:2000nf, Dong:2018cuv, Chandrasekaran:2022cip, Lin:2022nss} as well as probes of complexity such as \cite{Baiguera:2023tpt, Aguilar-Gutierrez:2024rka, Aguilar-Gutierrez:2023pnn, Aguilar-Gutierrez:2023zqm}.  Moreover, it may connect to approaches with the dual formulated on a spacelike boundary which also admit a formulation in terms of a $T^2$ deformation \cite{Araujo-Regado:2022gvw}. See e.g. \cite{Galante:2023uyf} for a review of some recent developments.}  After developing this theory in the main paper, we will comment on potential generalizations to 4d in the discussion section.   

To begin, let us summarize the structure of the universal part of the theory, captured in \cite{Coleman:2021nor}.  
A holographic seed CFT of central charge $c$ has two universal bands of energies:  the vacuum and a band of energies within order $c^0=1$ of $\Delta = c/6$ with a state count obeying the Cardy formula \cite{Hartman_2014, Mukhametzhanov:2019pzy} which matches the Bekenstein-Hawking entropy for the black hole of radius equal to the AdS curvature radius.  Deforming this theory by the $T\bar T+\Lambda_2$ deformation in the way defined in \cite{Coleman:2021nor}, one solves for the spectrum all along the deformation via the method discovered by \cite{Smirnov:2016lqw}.
For these universal bands,  the dressed energy \cite{Smirnov:2016lqw} of the solvable model as a function of the deformation parameter (see \eqref{eq-pure-gr-energy} for its explicit form)  matches the Brown-York quasilocal energy of the corresponding bounded patches of de Sitter as a function of the boundary size.\footnote{We should stress that the vacuum and $\Delta \simeq c/6$ energy bands are captured in a single theory (whereas in parts of \cite{Coleman:2021nor}, they were studied separately using different deformation trajectories).  They arise together in the same theory once we retain both signs of the square root in the dressed energy formula \eqref{eq-pure-gr-energy} and hence in the Hamiltonian.}    This precise map includes the contributions of the extrinsic curvature fluctuations known as `boundary gravitons', as worked out in detail for Dirichlet-bounded $AdS_3$ in \cite{Kraus:2022mnu}.\footnote{In particular, they reproduce on the gravity side a square root formula for energies, matching the $T\bar T$ deformation.  This cuts off the real spectrum at a finite energy level, giving a finite quantum system which does not contain divergent extrinsic curvature fluctuations. The authors \cite{Kraus:2022mnu} report that this reduces explicitly to \cite{Iliesiu:2020zld} in $2d$ (see  \cite{Stanford:2020qhm} for another interesting prescription for UV completion of the $2d$ problem).  It would be interesting to generalize the analysis of \cite{Kraus:2022mnu} to the $T\bar T+\Lambda_2$ phase, but the same general map between gravity equations and the deformation equations indicate a square root cutoff on the spectrum also in this case. We discuss this further in \S\ref{sec-grav-sector}.} 
Along the integrable deformation, one can follow the energy levels individually, with the count of states in the real dressed spectrum matching the horizon entropy.  The solution of the theory via the method discovered by \cite{Smirnov:2016lqw} is valid at finite central charge $c$.

\begin{figure}[t!]
  \centering
  \includegraphics[width=0.95\linewidth]{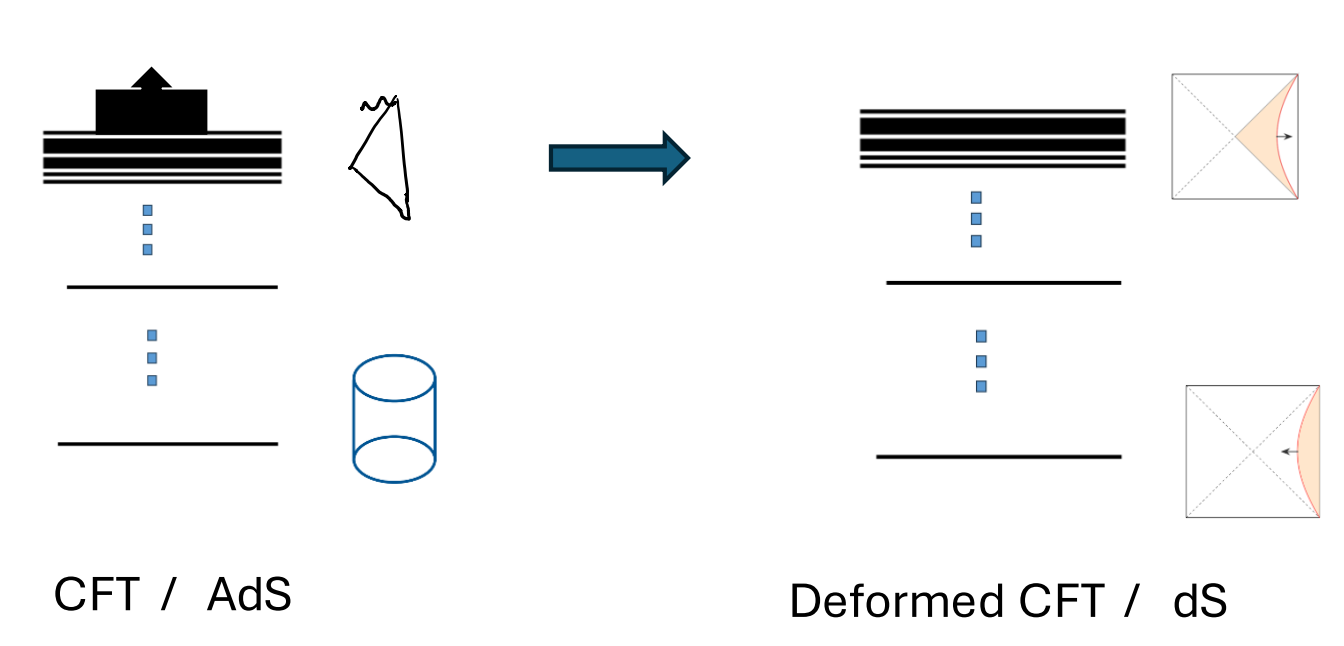}
  \caption{ \small \sffamily 
Schematic of the seed spectrum dual to the states of quantum gravity in $AdS_3$ deforming into the spectrum of the finite quantum theory dual to a bounded patch of dS.  Starting from a holographic conformal field theory and deforming according to a specific generalization of the $T\bar T$ deformation brings the system to a finite quantum system matching the entropy and quasilocal energy (hence radial geometry) of a bounded patch of de Sitter.  Previous work \cite{Coleman:2021nor} obtained this for the universal gravitational sector which contains the deformed $\Delta \simeq c/6$ band of energies and the vacuum state.  These are captured with the solvable $T\bar T+\Lambda_2$ deformation whose finite real spectrum explains the finite de Sitter entropy.  The present work generalizes the deformation to capture bulk matter, defining a finite quantum system whose leading approximation at large $c$ matches general relativity plus matter with local Dirichlet boundary conditions at the timelike boundary. In this system without fluctuating boundary gravity, the thermodynamic relations take the standard form for an ordinary quantum system \cite{Banihashemi:2022htw, Banihashemi:2022jys}. In addition to the depicted states, the theory admits domain wall configurations between AdS and dS.}
  \label{fig:universal-deformation-simpler}
\end{figure}

This universal and solvable deformation captures the gravitational sector of the theory, but not the model-dependent details of bulk matter, including the fine structure in the dressed $\Delta \simeq c/6$ energy band.  Aside from any energy level which precisely sits at $\Delta =c/6$,  states in this band of energies are not exactly continuous in the pure $T\bar T+\Lambda_2$ deformation\footnote{This occurs at the step where $\Lambda_2$ enters and the branch choice flips in the square root appearing in \eqref{eq-pure-gr-energy}. In the present work, full continuity will be restored via matter effects that capture the uplift from AdS to dS.}.  These discontinuities do not affect the solvability of the theory, as we can follow these subleading jumps in the energy and obtain an explicit finite real spectrum of energies $E_n$ which simply defines the Hamiltonian $H=\sum_{\text{levels}~ n} E_n |n\rangle \langle n|$ everywhere along the deformation.   
Happily the universal sector contains the most entropic energy band, whose fine structure is not needed for the state count (entropy) which matches the horizon area of black holes in AdS and the bare cosmic horizon in de Sitter.

The solvable theory captures the first two terms in the generalized entropy   
\begin{equation}\label{eq-generalized-entropy}
    {\cal S}_{\text{gen}} = {\cal A}/4G_N - 3 \log({\cal A}/4 G_N) + S_{\text{fields}};
\end{equation}
in terms of the horizon area ${\cal A}$,
along with the appropriate notion of energy defined in the presence of a timelike boundary by Brown and York \cite{Brown:1992br}.
It contains subleading contributions from bulk matter, which also imprints order $G_N^0$ contributions to the area term (via order $G_N$ back reaction on the area) as well as to the Brown-York energy.
With a timelike boundary with frozen gravity, general relativity plus effective quantum field theory (GR + EFT) obeys the standard first law of thermodynamics  
\begin{equation}
    \delta E_{\text{Brown-York}}=\delta E_{\text{bulk-matter}}+ T\delta {\cal S}
\end{equation}
discovered in \cite{Banihashemi:2022htw, Banihashemi:2022jys}.  This thermodynamics emerges from the statistical mechanics of the boundary theory we develop in this work, adding matter to the solvable $T\bar T+\Lambda_2$ deformation to produce an ordinary quantum system.

To capture the details of bulk matter fields, relevant both for fine structure of the dominant $\Delta \simeq c/6$ energy band and for less entropic energy bands of the system, we must generalize the formulation of the theory.  
The reason is well known \cite{Guica:2019nzm}\cite{Kraus:2018xrn}: the solvable $T\bar T(+\Lambda_2)$ theory is consistent with a bulk cutoff \cite{McGough:2016lol} (\cite{Gorbenko:2018oov, Lewkowycz:2019xse, Coleman:2021nor}) for gravity, but does not endow bulk quantum fields with local boundary conditions at the Dirichlet wall \cite{Kraus:2018xrn, Guica:2019nzm}.  One way to see this intuitively is that multitrace deformations \cite{Aharony:2001pa, Aharony:2001dp} change the boundary conditions of bulk fields \cite{Berkooz:2002ug, Witten:2001ua, Mueck:2002gm}; the pure $T\bar T + \Lambda_2$ deformation does this only for gravity.  A generalization to other bulk fields dual to operators ${\cal O}$ would require a generalization of the schematic form
\begin{equation}\label{eq:deformation-schematic-draft}
    \partial_\lambda S \sim \int T\bar T + {\cal OO} + \Lambda_2
\end{equation}
along with an explicit uplift from AdS to dS to account for the transition to nonzero $\Lambda_2$.  This has an immediate difficulty: adding such (generically irrelevant) composite operators leads to ambiguities, suggesting that such a construction would have meaningless output wholly dependent on input assumptions.   
At infinite $c$, composite operators factorize, enabling a clear treatment of \eqref{eq:deformation-schematic-draft} reverse-engineered from the gravity side to ensure local Dirichlet boundary conditions for all fields at the boundary (along the lines of Section 3.3 of \cite{Hartman:2018tkw}).  But that as it stands leaves open the question of the existence of a finite-c completion.

In this work, we resolve this, constructing a finite-$c$ completion and incorporating local bulk matter into the duality \cite{Coleman:2021nor}.  To do so, we formulate a finite system whose large central charge ($c\gg 1$) approximation reproduces the deformation required by the large-c factorization results and their implications for the dressed energies.  
The finiteness of our system is obtained by regulating our construction with an arbitrarily small initial pure $T\bar T$ deformation, giving a finite real spectrum.  Starting from this finite quantum system, we can define the operators required in the deformation \eqref{eq:deformation-schematic-draft} by implementing explicit subtractions of singular terms, so as to match the large-$c$ definition of the operators up to small corrections.  The theory is defined by this algorithm: at each step of the deformation we update the theory by recomputing $T, {\cal O}$ and add the appropriate composite operators \eqref{eq:deformation-schematic-draft} with the subtractions.  
In order to capture bulk processes involving not only low energy fields, but also Kaluza-Klein modes and other high energy excitations one needs operators corresponding to each. 
We will denote by $GR + EFT + \dots $ the degrees of freedom whose observables we will match to in our prescription for subtracting divergences and defining the renormalized operators, with the $\dots$ indicating elements of string/M theory that enter.
In our full trajectory, the internal space is never larger than the bounding cylinder, with scale separation in the de Sitter phase arising from the standard 3-term structure of moduli stabilization \cite{Dong:2010pm, DeLuca:2021pej}. 
On the other hand, to capture only processes accessible at low energies in the bulk, one can consider a more minimal version based on EFT without insisting on a particular string theory realization.
In either case, the theory thus defined serves to define the quantum gravity theory, which contains finite $c$ corrections beyond those contained in  $GR + EFT +\dots$.      

GR + EFT plus low energy string/M theory is not always valid; we match its predictions when it applies.  For example, during the uplift of the deformation, the low energy GR and EFT description breaks down for the states near $\Delta =c/6$:  in the gravity-side description, the boundary skirts the horizon, leading to strong fluctuations.\footnote{This was studied in \cite{Silverstein:2022dfj} for a Neumann boundary condition for compactification degrees of freedom in string/M theory.}
We can view this as a feature rather than a bug:  the theory is less constrained at this step. Still, in the bulk of our deformation we keep track of operators dual to bulk fields that become light in either the AdS or dS phase as well as a scalar $\Phi_u$ that interpolates between the AdS and dS vacua. For states well below $\Delta =c/6$, the uplift occurs within the regime of validity of $GR + EFT + \dots$ and we define it consistently with that simply by deforming the boundary conditions of $\Phi_u$ to interpolate from the value it takes in an AdS vacuum to a value it takes in a dS vacuum.\footnote{Explicit compactifications uplifting AdS/CFT to dS are available in 3 and 4 external dimensions \cite{Dong:2010pm, DeLuca:2021pej}.}  
This restores continuity of the model-dependent low-lying energy levels far from $\Delta \simeq c/6$ along the trajectory.

This subtraction criterion does not specify a unique theory; different versions of the theory which all agree with low energy general relativity and effective field theory in its regime of applicability differ in a way that is subleading at large $c$.  It may be that additional criteria are desirable, beyond simple matching the low energy theory obtained by deforming AdS/CFT dual pairs. 
We note, however, that the landscape of string vacua leads to a vast multiplicity of internal configurations consistent with similar low-energy physics, including infinite sequences of tractable models \cite{Delacretaz:2016nhw}.  We may view the non-uniqueness of our construction in a similar vein, though the origin of the multiplicity is different.

\begin{figure}[t!]
  \centering
  \includegraphics[width=0.95\linewidth]{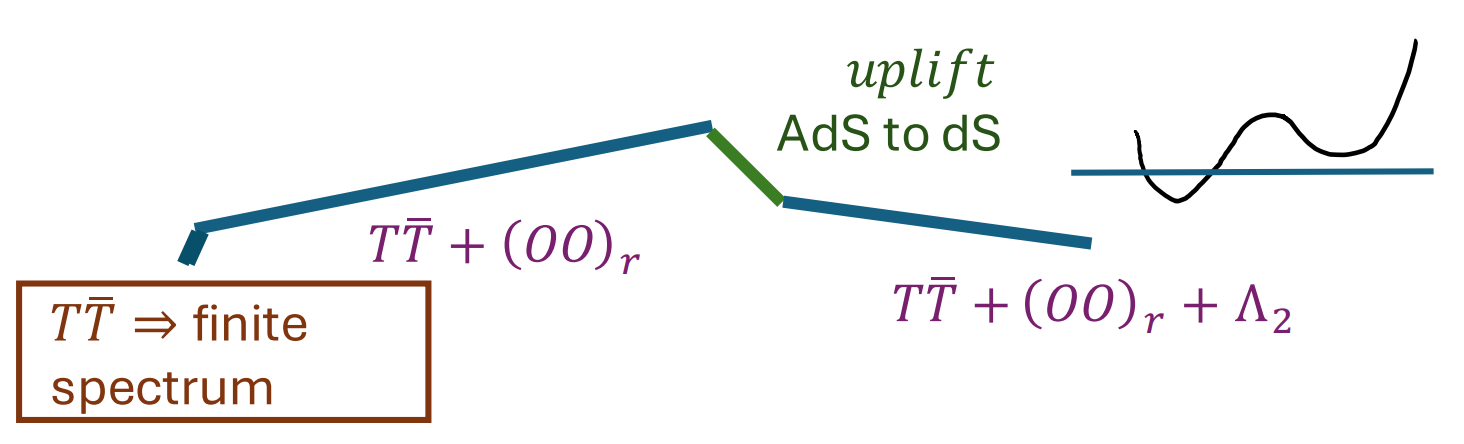}
  \caption{ \small \sffamily 
A schematic of our algorithm to define a holographic dual for the $dS_3$ static patch with timelike non-gravitational boundary.  The explicit cut off on the real spectrum  from a small initial $T\bar T$ deformation enables one to prescribe a concrete subtraction procedure to define the operators $(\mathcal{OO})_r$ required to endow bulk matter fields with approximately local boundary conditions.}
  \label{fig:trajectory-4-part-schematic}
\end{figure}

See Figure \ref{fig:trajectory-4-part-schematic} and Algorithm \ref{algo} for a summary of the defining algorithm for our theory.\footnote{We are implementing here a schematic first order explicit Euler integration, designed just to highlight the step-wise nature of the algorithm. We do not expect any particular problem with stability that would require a more stable method. However, this might be an important question if one wanted to use this algorithm to explicitly compute the deformed theory on a computer.}
The prescription agrees with $GR + EFT + \dots$  where they apply (including for the description of the dressed $\Delta \simeq c/6$ states far from the uplift step).  The strong quantum gravity effects arising from the details of the boundary, at the uplift for the dressed $\Delta \simeq c/6$ states, and for high energy processes in the bulk are defined by the non-gravitational boundary quantum mechanics theory.     
This construction is quite formal and rather indirect; it is defined by an algorithmic update rule and does not include a full resummation of the Hamiltonian.\footnote{See e.g. \cite{Susskind:2021esx, Narovlansky:2023lfz}} But it can be used to address key questions in de Sitter, in particular verifying the microstate-count interpretation of the refined \cite{Anninos:2020hfj} Gibbons-Hawking entropy \cite{Gibbons:1977mu} derived from the solvable $T\bar T+\Lambda_2$ core of our more general model.     

To use our boundary theory to illustrate the emergence of bulk locality, we generalize the calculation of dressed energy levels to capture time-dependent matter excitations $\langle {\cal O}(t)\rangle \sim \sin\omega t$ in the leading nontrivial order.  In general, this would require generalizing the steps of Smirnov and Zamolodchikov's derivation of the differential equation for dressed energies \cite{Smirnov:2016lqw} to incorporate general time dependent excitations. We can more simply analyze situations where the expectation value of the pressure in an energy eigenstate takes a simple form.
As we will see, at least in that case this independent boundary-side calculation can be done trivially and matches the physics of the gravity side, including bulk matter with local boundary conditions.  

Our analysis of the gravity-side solutions in general illustrate basic differences with respect to the pure gravitational case.  In particular it is important to note that the deformation cannot be described as moving a boundary through a pre-existing bulk solution.   Rather, the gravity-side counterpart to the dressed energy spectrum arises by keeping track of the solutions as a function of the bounding cylinder size $L_c$; for each $L_c$ there is a new space of solutions.  
In addition, the solution space of GR + EFT that we derive below will reveal another significant feature:  the effect of matter excitations on the dressed energy formula \eqref{eq-pure-gr-energy} appears multiplying the square root factor in that expression, thus not affecting the reality of the finite spectrum.

Another outcome is the determination that cosmic horizon patch holography has a type I Von Neumann operator algebra.  One can think of the boundary here as somewhat analogous to  the observer in \cite{Chandrasekaran:2022cip}, where the Newton constant was taken to zero at the end, engendering a type II algebra.  In our finite quantum system matching gravity-side physics at finite $G_N$, we obtain type I. 
We note that the effect of bulk matter on the generalized entropy \eqref{eq-generalized-entropy} is subleading to the $A/4 G_N$ term, contained in model-dependent contributions to the logarithmic term (as computed in \cite{Anninos:2020hfj} but with the Dirichlet boundary conditions on the wall).  As stressed above, the $T\bar T + \Lambda_2$ deformation captures the area term plus the universal part of the logarithmic correction \cite{Coleman:2021nor}; in this work we see this survives with matter included in a finite quantum system.

Altogether, this brute force but well-defined prescription provides a holographic dual for the large-radius $dS_3$ static patch with timelike boundary, with the correct emergent static patch geometry, entropy, and approximately local bulk matter dynamics.\footnote{It would be interesting to connect it to recent proposals for a concrete dual to dimensionally reduced $dS_3$ via the SYK model \cite{Narovlansky:2023lfz, Susskind:2021esx}.}

\begin{algorithm}[htb!]
   \caption{\small \emph{$T\bar{T}+\mathcal{O}{O}+uplift+\Lambda_2$}.
   $\Delta y, \epsilon_u$: step-sizes in theory space (default: infinitesimal).
   } 
   \label{algo}
  \begin{algorithmic}[1] 
\State \textbf{Require:} A 
holographic seed CFT of central charge $c$ living on a cylinder with size $L$, with its generating functional of correlation functions $\mathcal{W}$ and set of energy eigenstates $ \{\varphi^{(0)}_n\}$ with energies $\{E^{(0)}_n\}$, and large-N EFT + GR +\dots correlators (with $\dots$ including KK and string/M theory effects)
\State \textbf{Require:} A scalar field $\Phi_u$ with a potential $V$ interpolating from the holographic AdS vacuum at $\Phi_{uA}$ to a dS minimum at $\Phi_{uS}$, along with coupled fields $\Phi_\perp$.
\State \textbf{Require:} $y_T$ (explicit regulator).

\State $y \gets 0$ \Comment{Initialize dimensionless deformation parameter}
\State $y_0 \gets 3/c\pi^2$
\vspace{4pt}
\Function{TruncateSpectrum}{$\{E_n\}, \{\varphi_n\}$}
      \For{$E_i \in \{E_n\}$}
        \If{$E_i$ \texttt{is complex}}
            \State \texttt{remove} $(E_i, \varphi_i)$ \texttt{from} $(\{E_n\},\{\varphi_n\})$
        \EndIf
    \EndFor
 
    \State \Return $\{E_n\}, \{\varphi_n\}$
\EndFunction
\vspace{4pt}
\State $\Phi \gets \Phi_A$

\While{$y < y_T$}
\State $\{E_n\}, \{\varphi_n\} \gets \Call{TruncateSpectrum}{\{E_n\}, \{\varphi_n\}}$
     \State $T_{ij} \gets \frac{-2}{\sqrt{-\gamma}}\frac{\delta \mathcal{W}}{\delta \gamma^{ij}}$
    \State $\mathcal{W} \gets \mathcal{W} - L^2\frac{\pi}{4} \Delta y \int d^2x \sqrt{-\gamma} (T^{ij}T_{ij} - (T^i_i)^2)$
    \State $y \gets y + \Delta y$
\EndWhile
\vspace{4pt}

\While{$y > y_T$}
    \If{$ y_0 - \frac{\Delta y}{2} < y <  y_0 + \frac{\Delta y}{2} $}
    \Comment{Perform the uplift}

    \State $\sigma \gets 0$
    \While{$\sigma \leqslant 1$}
    \State $\Phi_u\gets \sigma \Phi_{uS} +(1-\sigma) \Phi_{uA}$
    \State $\sigma \gets \sigma + \epsilon_u $
    \EndWhile
    \EndIf
    \State $\{E_n\}, \{\varphi_n\} \gets \Call{TruncateSpectrum}{\{E_n\}, \{\varphi_n\}}$

    \State $T_{ij} \gets \frac{-2}{\sqrt{-\gamma}}\frac{\delta \mathcal{W}}{\delta \gamma^{ij}}$
    \State $\mathcal{O} \gets -\frac{\delta \mathcal{W}}{\delta J}$
    \State   $ \langle universal| (\mathcal{OO})_r \mathcal{O}_1\dots \mathcal{O}_m | universal\rangle \gets \langle universal | (\mathcal{OO})_r \mathcal{O}_1\dots \mathcal{O}_m | universal \rangle_{EFT+GR+\dots}$
    \State $\mathcal{W} \gets \mathcal{W} - L^2\frac{\pi}{4} \Delta y \int d^2x \sqrt{-\gamma} (T^{ij}T_{ij} - (T^i_i)^2 -\frac{1}{L^4 y^2}(1+\texttt{sign}(V(\Phi))+(\mathcal{O}\mathcal{O})_r)$
    \State $y \gets y + \Delta y$
\EndWhile
\end{algorithmic}

\end{algorithm}

\section{Comments on the gravitational sector}\label{sec-grav-sector}

Before introducing our framework for local bulk matter, we would like to clarify some aspects of the gravitational sector covered by the solvable $T\bar T + \Lambda_2$ deformation in \cite{Gorbenko:2018oov, Coleman:2021nor}. 

\subsection{The dressed classical action for $T\bar T+\Lambda_2$ and role of `boundary gravitons' (extrinsic curvature fluctuations)}

The key relation  underlying the duality at the pure gravity level is the following  \cite{McGough:2016lol, Gorbenko:2018oov}.  The $3d$ bulk Einstein equations in the presence of a Dirichlet boundary, where the intrinsic geometry of the boundary is fixed, imply the defining equation of the $T\bar T + \Lambda_2$ deformation  
\begin{equation}\label{eq-EE-to-deformation}
    T^i_i = 4\pi\lambda T\bar T -\frac{1}{\pi\lambda}(1-\eta) \Rightarrow \partial_\lambda S = -\int [2\pi T\bar T - \frac{1}{2\pi\lambda^2}(1-\eta) ]
\end{equation}
where in the gravity-side variables, $T$ is the Brown-York stress tensor \cite{Brown:1992br}, and where $T\bar T = -\frac{1}{4}\det(T)$.  This 
indicates that pure gravitational configurations in the Dirichlet problem in $3d$ are governed by the $T\bar T +\Lambda_2$ deformed theory.\footnote{We would like to remark that the initial boundary value problem, including geometric uniqueness, is not in conflict with special cases of the Dirichlet condition (such as a cylinder boundary) even in four dimensions, see e.g. section 5.5 of \cite{Anninos:2023epi}.}  

Among the $T\bar T+\Lambda_2$ dressed states of the CFT are boundary graviton excitations.  Since these figure into some of the analyses of subtleties with Dirichlet boundaries \cite{Marolf:2012dr, Witten:2018lgb} (see also \cite{Stanford:2020qhm, Iliesiu:2020zld} in 2d), it is worth spelling out how they fit in.\footnote{We thank H. Maxfield and D. Stanford for discussions of this point. }     The boundary gravitons are extrinsic curvature fluctuations arising from deformations of the embedding of the boundary, consistently with the Dirichlet condition that it have fixed intrinsic geometry.  This scalar degree of freedom $\phi_g$ was studied at the linearized level in bounded patches of $AdS_3$ e.g. in \cite{Marolf:2012dr} and nonlinearly in \cite{Kraus:2022mnu, Ebert:2022cle}.  The latter works provide extensive evidence directly on the gravity side that this degree of freedom is governed classically by the Born-Infeld Lagrangian, which is indeed the $T\bar T$ dressed theory of a free boson.  Generalizing the calculation of the dressed classical action for a free boson to include the $T\bar T + \Lambda_2$ deformation, e.g. using the elegant prescription in \cite{Bonelli:2018kik}, yields
\begin{equation}
     {\cal L} =  -\frac{1}{\pi\lambda}\left(1-\sqrt{\eta - \pi\lambda \partial_\mu\phi_g\partial^\mu \phi_g} \right)
\end{equation}
which satisfies the deformation equation \eqref{eq-EE-to-deformation}. 
Moreover, \cite{Kraus:2022mnu} report that the theory they derive reduces explicitly to a version of cut off 2d JT gravity \cite{Iliesiu:2020zld} (see   \cite{Stanford:2020qhm} for another interesting prescription for UV completion of $2d$ JT gravity with finite renormalized boundary).

Linearized boundary graviton dynamics \cite{Marolf:2012dr} arises from expanding this action around a suitable background configuration \cite{Cooper:2013ffa}.  In the full nonlinear quantum system,
since the real spectrum of the $T\bar T+\Lambda_2$ deformed theory is finite, and matches to the Dirichlet problem in $3d$ gravity, there is no buildup of arbitrarily strong extrinsic curvature fluctuations.

\subsection{A single theory captures all the universal energy levels}

Universal states of bounded de Sitter quantum gravity include the cosmic horizon patch and the pole patch, as depicted on the right side of figure \ref{fig:universal-deformation-simpler}.  In this subsection, we would like to clarify that the version of the $T\bar T + \Lambda_2$ trajectory defined in \cite{Coleman:2021nor} -- which we briefly review here -- can capture in one theory both the quantum state corresponding to the cosmic horizon patch and the pole patch. 

This trajectory is defined by \eqref{eq-EE-to-deformation}, proceeding in two segments.  First is evolution by pure $T\bar T$ with $\eta=1$ until a value $y=y_0$, after which one turns on the $\Lambda_2$ part with $\eta=-1$.  (In the present work, we will resolve this step with a matter sector modeling the uplift from AdS to dS.)   With $y_0=3/c\pi^2$, the resulting dressed energy formula is \cite{Coleman:2021nor}
\begin{equation}\label{eq-dressed-En-review-pure-GR-sec}
    EL = {\cal E}=\frac{1}{\pi y}\left(1\mp\sqrt{\eta+\frac{y}{y_0}(1-\eta)-4\pi^2 y\left(\Delta-\frac{c}{12}\right)+4\pi^4 y^2J^2} \right)\,,
\end{equation}
where $\Delta$ is the dimension corresponding to each seed CFT energy level that is being dressed.  

We have indicated the possibility of states with both signs of the square root here.  Indeed, the quasilocal energies of the cosmic horizon patch and the pole patch fit the upper and lower signs of \eqref{eq-dressed-En-review-pure-GR-sec} with $\Delta = c/6$.  We may construct the Hamiltonian $H=\sum_n E_n |n\rangle\langle n|$ keeping states with both signs of the square root. 
This fits with the gravity side:  fixing the proper geometry of the boundary according to the Dirichlet boundary condition does not constrain the extrinsic curvature $K$, which may take either sign. As explained in \cite{Coleman:2021nor} and reviewed below, this corresponds to the two branches of the square root in \eqref{eq-dressed-En-review-pure-GR-sec}.  

This does not imply that there are equal numbers of states in the two sectors corresponding to the two branches of the square root.  In the gravity description, one determines nonsingular solutions obeying the boundary conditions, whose existence and properties can depend on the sign of $K$. In the dual deformed-CFT description, in constructing the Hamiltonian at each step in the trajectory, the number of states with each sign of the square root can be different.  
At $y=y_0=3/c\pi^2$, where we turn on $\Lambda_2$, this works as follows.   Constructing the pole patch from the $-$ sign in \eqref{eq-dressed-En-review-pure-GR-sec}, we include one such state,\footnote{The small entropy of the pole patch, whose boundary excludes the cosmic horizon, is independently checked by a different $T\bar T+\Lambda_2$ trajectory with $y_0\to\infty$ \cite{Coleman:2021nor, Gorbenko:2018oov}.} while for the cosmic horizon patch we keep all of the $\Delta \simeq c/6$ states comprising the horizon microstates.  For the latter states, the gravity description at $y=y_0$ consists of a boundary skirting the horizon, for which the AdS BTZ black hole and the dS cosmic horizon are indistinguishable.\footnote{An argument for this persisting in string theory, with its internal degrees of freedom from compactification, appears in \cite{Silverstein:2022dfj}}

So far we have obtained both the cosmic horizon (CH) patch and the pole patch from the $\Delta\simeq c/6$ part of the energy spectrum. We must also address the fate of the $CFT$ vacuum state ($\Delta =0$) along our trajectory.  This yields a state with a gravity-side description involving a domain wall:  as $y$ approaches $y_0=3/c\pi^2$ where the uplift will occur, the Dirichlet wall on the gravity side still bounds a large region of AdS, distinguishable from dS.  Continuing (increasing $y$) after adding $\Lambda_2$ pulls out the boundary, changing the quasilocal energy according to the radial evolution in dS.

\section{Defining the bulk-local deformation as a finite system:  type I operator algebra}

In this section we will build up the definition of our deformed theory.  First we will lay it out at large $c$, refining the treatment in section 3.3 of \cite{Hartman:2018tkw}.  Then in \S\ref{sec-finiteness} we will explain how to adjust the construction to obtain a well defined system at finite $c$.  

Our construction will generalize the duality \cite{Coleman:2021nor} between solvable models and pure 3d gravity quantities.  The precise dictionary between bulk and boundary theory parameters derived in \cite{McGough:2016lol, Gorbenko:2018oov} and applied in \cite{Coleman:2021nor} will still apply in our theory.\footnote{The construction \cite{Coleman:2021nor} in turn realized and explained the `numerology' in \cite{Bousso_2002}.}  
The seed CFT has large but finite central charge $c$.  The deformed CFT  lives on a cylinder of fixed spatial size $L$, evolving along a trajectory parameterized by the dimensionful parameter $\lambda$.  
It is useful to define a dimensionless deformation parameter
\begin{equation}\label{eq-y-defs}
    y=\frac{\lambda}{L^2} = \frac{8 G_N\ell}{L_c^2}
\end{equation}
where the last expression uses
the dictionary to the gravity side 
\begin{equation}\label{eq:param-dictionary-matter}
    c=\frac{3\ell}{2 G_N}, ~~~ \lambda=8 G_N\ell \frac{L^2}{L_c^2}\,.
\end{equation}
Here $\ell$ is the curvature radius of the AdS or dS that we start and end with along the trajectory we will define.  

Along our trajectory, $L, c$ and $G_N, \ell$ are fixed on the boundary theory and gravity side respectively. On the boundary theory side $\lambda$ evolves along the trajectory, and on the gravity side the proper size $L_c$ of the bounding cylinder changes along the trajectory.  These agree on the evolution of $y$ \eqref{eq-y-defs}.

\subsection{Large $c$ construction}\label{sec:large-c}

In this subsection, we will derive the necessary generalization of $T\bar T+\Lambda_2$ to incorporate local bulk matter, at the large-c level.  This by itself is not a complete theory, but it will be a good approximation to the finite-$c$ theory we develop in the subsequent sections.   

As described in \cite{Aharony:2018vux}, in the large $c$ limit we fix $\lambda c$.  At a given step in the deformation, we will add contributions including the double trace terms 
\be
\delta {\cal L} = 2\pi \delta\lambda T\bar T + f_O(c\lambda) {\cal OO}\,.
\ee
The first coefficient here is of order $1/c$, matching the prescription for multitrace deformations for matrix-like theories introduced in \cite{Aharony:2001pa}; in our bulk $3d$ context an example of such a seed theory is the CFT obtained from the low energy limit of the D1-D5 system.  In general we treat the operators ${\cal O}$ dual to bulk matter fields according to their particular representations (which can include for example composites of singlets and fundamentals in addition to traces of products of adjoints and bifundamentals). We will consider to be specific bulk matter fields whose internal dynamics is independent of $G_N$ and will specify below our convention for the definition of these operators.

Following \cite{McGough:2016lol, Kraus:2018xrn, Gorbenko:2018oov, Hartman:2018tkw}
we work from the Einstein equations, including  matter stress-energy.  The type of matter that is most relevant is (i) bulk gauge field sectors, which are naturally light, and (ii) a sector effecting the uplift from negative to positive bulk cosmological constant.  Since in the three dimensional bulk we can dualize vector fields to scalars, we will write everything in terms of scalars.

Let us start by deriving a formula for the deformation including a bulk scalar field with stress-energy tensor ${\cal T}$.  We will work with a patch of spacetime bounded by a cylinder of proper size $L_c$. We can coordinatize this with metric  
\begin{equation}
    ds^2 = dw^2 + g_{ij}(w, x) dx^i dx^j
\end{equation}
where we impose boundary conditions on $g_{ij}$ such that there is a boundary cylinder of size $L_c$.  Without loss of generality we denote its radial position as $w=w_c$.  
The three-dimensional bulk action is
\be\label{eq:action0}
S= \int_{\mathcal M} d^3x\,\sqrt{-g} \left[ \frac{1}{16\pi G_N} R^{(3)}-\frac{1}{2}(\partial \Phi)^2-V(\Phi)\right]+ \frac{1}{8\pi G_N} \int_{\partial \mathcal M} d^2x\,\sqrt{-g} \left(K-\frac{B_{CT}(\Phi)}{\ell} \right) \,,
\ee
with metric signature $(-++)$. For a pure AdS solution, $V=-2/\ell^2 G_N$ and $B_{CT}=1$, but we allow for a more general matter potential. The boundary term proportional to the trace of the extrinsic curvature is needed in order to have a well-defined Dirichlet variational principle; we have also allowed for a boundary conterterm action determined by the function $B_{CT}(\Phi)$.\footnote{There is much freedom in the counterterm action, which must cancel all divergences at the beginning of the trajectory. It is also possible to add counterterms containing derivatives, such as $R^{(2)}$ and $(\partial \phi)^2$, but these will not play an important role in our discussion; so we will not consider them.}

We denote the bulk matter stress tensor by
\be
{\mc T}_{\mu\nu}=- \frac{2}{\sqrt{-g}} \frac{\delta S_{\text{matter}}}{\delta g^{\mu \nu}}\,,
\ee
and the Einstein tensor 
\be
E_{\mu\nu}=\frac{1}{8\pi G_N}\left(R_{\mu\nu}- \frac{1}{2} g_{\mu\nu} R\right)\,.
\ee

The radial-radial Einstein equation $E^w_w= {\cal T}^w_w$ combined with the definition of the Brown-York stress energy tensor,
\be\label{eq:Tdef}
T_{ij} =-\frac{2}{\sqrt{-g}} \frac{\delta S_{\text{on-shell}}}{\delta g^{ij}} = -\frac{1}{8\pi G_N} \left(K_{ij}-g_{ij}K+g_{ij} \frac{B_{CT}}{\ell} \right)
\ee
yield   
\be\label{eq:TFE-two-vacua}
T^i_i=\frac{1}{B_{CT}(\Phi)}\left[
4\pi G_N\ell \left(T_{ij}^2-(T^i_i)^2\right) +\frac{\ell}{2}(\partial_w \Phi)^2 -\frac{\ell}{2} \partial_i\Phi \partial^i \Phi - \ell V -\frac{1}{8\pi G_N\ell} B_{CT}(\Phi)^2\right]
\ee
at the boundary $w=w_c$.  
\footnote{This is essentially equation (A.6) of \cite{Gorbenko:2018oov} with $\eta$ in the $\Lambda_2$ term replaced by 
$\eta \to -V + \frac{1}{2}(-\partial_i\Phi\partial^i\Phi+{\Phi'}^2)$, which is the combination that appears in the bulk stress-energy tensor component ${\cal T}^w_w$. Some terms have different signs from \cite{Gorbenko:2018oov}, because we define the Brown-York stress tensor as in \eqref{eq:Tdef}.} 

Next, we apply the $E^w_j$ Einstein equation 
\begin{equation}
    E^w_j =\frac{1}{8\pi G_N} \nabla^i(K_{ij}-Kg_{ij}) = {\cal T}^w_j=\partial_w\Phi\partial_j\Phi\,.
\end{equation}
Using (\ref{eq:Tdef}) to trade $K_{ij}$ for $T_{ij}$, this becomes
\begin{equation}
 \nabla^i T_{ij}=-\left(\partial_w\Phi + \frac{1}{8\pi G \ell}\partial_\Phi B_{CT}(\Phi) \right)\partial_j\Phi \,.
\end{equation}
As expected, $x$-dependent sources (i.e. $\partial_j \Phi \neq 0$), generically break space-time translations and lead to a non-conserved $T_{ij}$. Here we will be interested in a conserved energy momentum tensor, obtained for boundary conditions that can be either local Dirichlet ($\Phi=\Phi_0$) or local Neumann-like ($\partial_w\Phi +\partial_\Phi B_{CT}(\Phi)=0$).  
In the Dirichlet case, for example, we see that the $-\partial_i\Phi\partial^i\Phi$ term in the trace flow equation vanishes.
Let us focus to be specific on the Dirichlet boundary condition for all fields, though we see no obstruction to extending our results readily to other choices of boundary conditions.

The next step is to identify the boundary theory operator corresponding to the bulk scalar field.  This proceeds as in e.g. \cite{Hartman:2018tkw, Lewkowycz:2019xse}.  The natural generalization of the identification of $T_{ij}$ with the extrinsic curvature \cite{Brown:1992br} is the identification of ${\cal O}$ with the radial field momentum $\Pi$ corresponding to $\Phi$:
\begin{equation}\label{eq:pi-o-scaling}
\mathcal{O} = \ell^{-\Delta+3/2}\left(\frac{L_c}{L}\right)^\Delta \Pi = \left( \frac{\lambda c}{12}\right)^{-\Delta/2}\ell^{3/2}\Pi\,.
\end{equation} 
{ In AdS/CFT, this normalization ensures that $\langle \mathcal{O}(x)\mathcal{O}(0)\rangle \sim 1/|x|^{2\Delta}$.\footnote{Intuitively, the power of $(L_c/L)^\Delta$ cancels a power of the radial cutoff that appears in the gravity calculation, while the power of $\ell$ ensures that the engineering dimensions match.} Here we have chosen the same definition for the deformed theory, which gives a $c$-independent 2-point function. We note that this normalization differs from the one used in the double-trace deformation literature, where single-trace operators are $N$ times a trace of products of adjoint fields. These single-trace operators have 2-point functions of order $N^2$, as opposed to the $N$-independent normalization we use in this work. 

Although we have written for brevity a single scalar field and its dual operator, we will be interested in including two matter sectors. First, in our construction we will require a scalar sector encoding the uplift from negative to positive bulk cosmological constant.  In \S\ref{sec:uplift}\ we will show that this ensures continuity of the energy levels, including for those below the dominant $\Delta \simeq c/6$ states. Given our choice of normalization with a $G_N$-independent bulk kinetic term, the $1/G_N$ effects from the uplift sector $\Phi_u$ are encoded in a nonlinear potential of the form
\be
V(\Phi_u)= \frac{1}{G_N} \tilde V(\sqrt{G_N} \Phi_u)\,.
\ee 
Second, we will include a probe sector that contributes at $O(G_N^0)$, in order to capture local dynamical bulk fields in the bulk effective field theory.  Such fields arise naturally if their mass is protected to be $\ll 1/G_N$ (the bulk Planck scale).  Examples include gauge theory sectors.  In three dimensions, such gauge fields $A$ can be dualized to scalars $\phi$, $dA = * d\phi$.  
A Dirichlet boundary condition for the scalars corresponds to a Neumann one for the gauge fields and vice versa.\footnote{An advantage of the gauge theory sectors is the simplicity of identifying the operator in a simple way in the boundary dual.  It is obtained via the Noether current ${\cal J}$ for the corresponding global symmetry in the dual. An ensemble with fluctuating $\pi_\phi = \partial_w\phi+\dots$ as we layed out above corresponds to fixed rather than fluctuating ${\cal J}$, so it is obtained by a Legendre transform.} The interaction potential in the probe sector can in general be nontrivial; but for that sector we will find it convenient to illustrate our main points in terms of a free bulk scalar field { (with the aforementioned uplift sector illustrating nonlinear matter).}}

In order to compute the conjugate momentum associated to the radial evolution of the scalar field, we 
note that the counterterm in the action (\ref{eq:action0}) can be written as a bulk total derivative,
\be
-\frac{1}{8\pi G_N \ell}  \int_{\partial \mathcal M} d^2x\,\sqrt{-g}\,B_{CT}=-\frac{1}{8\pi G_N \ell} \,\int_{\mathcal M} d^2x dw \sqrt{-g} \left(\frac{1}{2} g^{ij}\partial_w g_{ij}\,B_{CT}+ \left(\partial_\Phi B_{CT}\right) \partial_w \Phi \right)\,.
\ee
The first term here, proportional to $\partial_w g_{ij}$, is responsible for the $B_{CT}$ contribution to the stress tensor. The second term, on the other hand, contributes to the scalar canonical momentum,
\be\label{eq:Pidef}
\Pi=\frac{1}{\sqrt{-g}} \frac{\delta S}{\delta(\partial_w \Phi)}=- \partial_w \Phi- \frac{1}{8 \pi G \ell} \partial_\Phi B_{CT}\,.
\ee
For our Dirichlet boundary condition, the conjugate momentum $\Pi$ is free to fluctuate, as befits a dynamical operator in the boundary theory. 

Variations of the boundary value $\Phi(w_c) \to \Phi_c+\delta\Phi_c$ of the scalar can be identified with a source $J\propto \delta \Phi_c$. A calculation similar to that in (\ref{eq:pi-o-scaling}) gives
\be
J=\left(\frac{\lambda c}{12}\right)^{\Delta/2-1} \ell^{1/2} \delta \Phi_c\,.
\ee
The source vanishes, $J=0$, for our self-contained system at fixed energy at any point along the deformation trajectory including during the uplift.  But we keep track of it because it enters into the derivation of the operator ${\cal O}$ required to move along the trajectory, adding the new ${\cal OO}$ at each step.     
In the boundary theory language, the operator ${\cal OO}$ that we need to add is obtained from $\langle {\cal O}\dots {\cal O}\rangle = \frac{\delta}{\delta J}\dots \frac{\delta}{\delta J} \log Z_{dresssed}$.  
In these variables, the trace flow equation becomes
\bea\label{trfl4}
T^i_i&=& \frac{1}{B_{CT}} \left[
4\pi G_N\ell \left(T_{ij}^2-(T^i_i)^2\right)+ \frac{\ell}{2}\Pi^2 \right] +\frac{1}{8\pi G_N} \frac{B_{CT}'}{B_{CT}}\Pi\nonumber\\
&&-\frac{\ell}{2B_{CT}} \partial_i\Phi \partial^i \Phi -\frac{\ell}{B_{CT}} V + \frac{1}{B_{CT}} \left(\frac{\ell}{2}\left(\frac{B_{CT}'}{8\pi G\ell}\right)^2-\frac{B_{CT}^2}{8\pi G\ell} \right)\,.
\eea

Lets focus on the term proportional to $B_{CT}' \Pi $ in the above expression, which is linear in $\mathcal O$. Perturbing the action by $\int d^2x\,\sqrt{-\gamma} J \mathcal O$ (with $\gamma$ the 2d metric) gives rise at first order in $\mathcal O$ to
\be
\int d^2 x \sqrt{-\gamma}  T^i_i \supset -\int d^2 x \sqrt{-\gamma} \beta(J) \mathcal{O} \,,
\ee 
with $\beta(J) = L_c\partial_{L_c}J$ the QFT beta function for $J$. Thus, comparing with (\ref{trfl4}), we read off the beta function
\be \label{eq:beta_J_holo}
\beta(J)= L_c\frac{\partial J}{\partial_{L_c}} =-\frac{1}{\pi} \left(\frac{\lambda c}{12}\right)^{\Delta-2} \frac{\partial_J B_{CT}(\Phi_{c} + J)}{B_{CT}(\Phi_{c} + J)}\,.
\ee
In an asymptotically AdS region, and choosing $B_{CT}$ as the standard holographic counterterm, this reproduces the correct holographic beta function (see e.g. \cite{deBoer:1999tgo}).

On the other hand, defining $\mathcal{W}[w_c,J]$ as the generating functional of correlation functions in the dual QFT at finite cutoff, we know that
\be
\int d^2 x \sqrt{-g} T^i_i = L_c \frac{d\mathcal{W}}{dL_c}= L_c\frac{\partial\mathcal{W}}{\partial L_c} + \int d^2 x \sqrt{-\gamma} L_c \partial_{L_c} J \frac{\delta\cal{W}}{\delta J}= L_c\frac{\partial\mathcal{W}}{\partial L_c} - \int d^2 x \sqrt{-\gamma} \beta(J) \mathcal{O} 
\ee 
where we made use of $\frac{\delta\mathcal{W}}{\delta J}=-\cal{O}$. Therefore we conclude that terms proportional to $\beta(J)$ cancel.
So far our discussion pertained to any scalar field.  We will illustrate our prescription with two scalar sectors:  a nonlinear uplift sector scalar $\Phi_u$ and a second, Gaussian scalar sector dual to the operator ${\cal O}$.
Writing $L_c\partial_{L_c} \mathcal{W} = -2 \lambda\partial_\lambda \mathcal{W}$, we end up with
\be
\begin{aligned} \label{floweq}
\lambda\frac{\partial \mathcal{W}}{\partial \lambda} &=\int d^2 x \sqrt{-\gamma} \frac{-1}{B_{CT}(\Phi_{uc}+J_u)} \left[2\pi \lambda T\bar{T} + \frac{1}{4}\left(\frac{\lambda c}{12}\right)^{\Delta-1}\left[\mathcal{OO}+\mathcal{O}_u \mathcal{O}_u\right]
  -\frac{1}{4}\left(\frac{\lambda c}{12}\right)^{2-\Delta}(\partial_i J)^2 \right.\\
& \left. - \left(\frac{\lambda c}{48}\right)^{1/2}\left(\frac{L_c}{L}\right)^3 V(\Phi_{uc} + J_u) + \frac{1}{4\pi^2 \lambda^2} \left(\frac{\lambda c}{12}\right)^{\Delta-1} B_{CT}'(\Phi_{uc}+J_u) - \frac{B_{CT}(\Phi_{uc}+J_u)^2}{2\pi \lambda}\right]\,
\end{aligned}
\ee
where the derivative in $B_{CT}'$ is taken with respect to $J_u$.\footnote{The third and the last term in \eqref{floweq} contain $L_c$ and $\Phi_{uc}$, but they can be expressed purely in terms of boundary quantities upon substituting explicit expressions for $V(\Phi)$ and $B_{CT}(\Phi)$. For example, in the absence of sources $J_u=0$, when the potential $V(\Phi_{uc})$ is at its AdS/dS minimum, they reduce to the $\Lambda_2$ term, $(\eta-B_{CT})/2\pi \lambda$.}

Below we will analyze in more detail the definition of the composite $\mathcal {O O}$ operator in the field theory side. Besides the semiclassical factorized contribution $\sim (\langle \mathcal O \rangle)^2 $, there are additional terms controlled by short-distance correlators. Our dual theory will turn out to be a finite quantum-mechanical theory, where these contributions are well-defined. Given this, we will define the  $\mathcal {O O}$ contribution to the trace flow equation by subtracting these effects; see the discussion below \eqref{eq-OO-finite-c-expansion-coefficient-interaction}.

We work with vanishing sources ($J=0$) to construct a self-contained bounded patch of spacetime. Putting all this together and looking ahead, we will arrive at a definition for the dual theory of the form of an explicitly renormalized version of  \eqref{eq:deformation-schematic-draft}.   Its existence is aimed first of all to validate the entropy count in \cite{Coleman:2021nor}.  Moreover, in \S\ref{sec-dressed-Es} we will show how to use it to compute some simple quantities of interest in the boundary theory and compare them to the gravity side.

\subsection{Finite system at finite $c$}\label{sec-finiteness}

As stressed in \S\ref{sec:intro}, the addition of the ${\cal O O}$ terms in our defining equation introduces new complications in the finite-$c$ setting of interest.  In order to obtain a well-defined theory, we will adjust the above prescription as follows.  The resulting theory is not as easily solvable as the pure $T\bar T+\Lambda_2$ theory, but it is well defined and we will see below that in simple cases we can generalize the dressed energy calculation to the leading nontrivial order in the effects of matter.  

Starting from the seed CFT, we first carry out a pure $T\bar T$ deformation to a very small value of the dimensionless deformation parameter $y=\lambda/L^2$; let us denote it $y_T\ll 1$.  At this point, the real spectrum is finite \cite{Smirnov:2016lqw}.  From this point on we embark on the $T\bar T + {\cal OO}+ \Lambda_2$ deformation.  

Of course, the reason we include the ${\cal OO}$ terms in the first place is to ensure bulk locality at the level of approximation corresponding to low energy bulk EFT.  The reason this is consistent with our initial pure $T\bar T$ trajectory is as follows.  We are free to take $y_T$ small enough that the initial violation of locality introduced by the pure $T\bar T$ deformation from $y$ to $y_T$ is only applicable to such high energy excitations that quantum gravity effects enter (for which there is no expectation of locality \cite{Polchinski:1998rq, Polchinski:1998rr}).   
Related to this, we stress that only infinite-energy excitations actually reach the original AdS boundary in AdS/CFT.  This includes excitations of massless fields for which the wave dynamics beyond the geometric optics approximation is important; a finite-energy excitation does not just follow the null geodesic trajectory (which does reach the boundary).  Our finite theory may be taken as the definition of the bulk quantum gravity, which appropriately deviates from EFT at such high energy scales.

Once the theory has a finite-dimensional Hilbert space, the additions along the main trajectory, including the ${\cal OO}$ terms, are all well defined: in a basis of energy eigenstates the (unrenormalized) operator is
\begin{equation}\label{eq-OO-E-basis-finite}
    \langle n|{\cal OO}|m\rangle = \sum_p \langle n|{\cal O}| p \rangle \langle p| {\cal O}|m\rangle 
\end{equation}
which is unambiguous since the sum over $p$ is finite.
In general the expectation value $\langle n| {\cal OO}|n\rangle $ does not factorize, in contrast to the special case of $T\bar T$ which factorizes up to a total derivative.
{Moreover, formally subleading contributions in the expansion around large $N$ can diverge as $y_T\to 0$.
Below in \S\ref{sec:boundary-prescription}, we will estimate the corrections to factorization and define an appropriate renormalized $\mathcal{OO}$ operator to add to our theory at each step in our algorithm.
}

After our $T\bar T$ deformation up to the small value $y_T$ of the dimensionless deformation parameter $y$, the dressed energies take the form
\begin{equation}
    {\cal E}_T = \frac{1}{\pi y_T}\left(1-\sqrt{1-2\pi y_T {\cal E}_{(0)}+ 4\pi^4 J^2 y_T^2} \right)
\end{equation}
First, we note that with $J=0$, the cut off on the real spectrum arises for seed dimensionless energy ${\cal E}_{(0)} = 2\pi(\Delta -c/12)$ of a maximal value ${\cal E}_{(0), max} = 1/2\pi y_T$. 
In fact, this scaling with $y_T$ persists in the presence of spin, as we can see in the following way as a consequence of the unitarity bound.

We would like to determine whether there is an upper cutoff on $\varepsilon_{(0)}$ such that for higher energies $\varepsilon$ becomes complex. This is controlled by
\be\label{eq:D0}
D_0=1-2\pi  \varepsilon_{(0)}\,y_T+4\pi^4 J^2\, y_T^2\,.
\ee
At a fixed $y_T$, one could think about trying to increase the spin $J$ such that $D_0>0$. However, the unitarity bound in $d=2$ is
\be
J \le \Delta\;\Rightarrow\;J \le \frac{1}{2\pi}\left(\varepsilon_{(0)}+ \frac{c}{12} \right)\,.
\ee
We are interested in the universal heavy states, which have $\varepsilon_{(0)} \sim c$, and we want to see whether we can make $\varepsilon_{(0)}$ arbitrarily large, at the same time increasing $J$, such that $\varepsilon$ stays real in the window of interest for $y_T$. So let us set
\be
J = \alpha \varepsilon_{(0)}\,,
\ee
and at sufficiently high energies unitarity requires
\be\label{eq:unitarity}
\alpha \le \frac{1}{2\pi}\,.
\ee
Returning to (\ref{eq:D0}), we have
\be
D_0=1-2\pi y_T\, \varepsilon_{(0)}+4\pi^4 y_T^2\,\alpha^2\,\varepsilon_{(0)}^2\,.
\ee
So $D_0$ has two roots as a function of $\varepsilon_{(0)}$,
\be
\varepsilon_{(0)}^\pm= \frac{1}{4\pi^3 \alpha^2 y_T} \left( 1 \pm \sqrt{1-(2\pi \alpha)^2} \right)\,.
\ee
For $\varepsilon \to 0$, $D_0>0$. If the two roots are complex, this means that $D_0$ will always stay positive for all $\varepsilon_{(0)}$, and then we could go to arbitrarily large $\varepsilon_0$ without the dressed energy becoming complex. However, because of the unitarity bound (\ref{eq:unitarity}) on $\alpha$, we see that $D_0=0$ always has two real roots. We conclude that for a given choice of spin compatible with unitarity, we need to impose a cutoff
\be\label{eq-energy-max-including=spin}
\varepsilon_{(0)} < \varepsilon_{(0)}^-= \frac{1}{4\pi^3 \alpha^2 y_T} \left( 1 - \sqrt{1-(2\pi \alpha)^2} \right) \equiv \frac{c_\alpha}{y_T}
\ee
so that the dressed energy stays real.
From \eqref{eq-energy-max-including=spin} we see that including spatial momentum $J$, the spectrum cuts off at a dimensionless energy $\propto 1/y_T$.

Let us now apply this to the question of the contributions to (\ref{eq-OO-E-basis-finite}).  
We note that although there are $e^{{\cal S}}$ states appearing in the spectral decomposition \eqref{eq-OO-E-basis-finite}, almost all of them have negligible form factors $\langle n | {\cal O} | p\rangle$.  
The usual UV ambiguities in defining the irrelevant deformation arise from the high momenta entering into the spectral decomposition of ${\cal OO}$, corresponding to a dimensionless energy ${\cal E}_p\sim |p|L$.
For an operator in the seed CFT of dimension $\Delta$, this goes like
\begin{equation}
    \int^{M_* } d^2 p(|p| L)^{2\Delta -2} \sim \left(\frac{c_\alpha}{y_T L}\right)^{2\Delta}\,.
\end{equation}
where $M_*$ is a UV cutoff scale, which for us is proportional to $\frac{1}{y_T L}$ right after the first, pure $T\bar T$ segment of our 4-part trajectory.   
In an energy eigenstate, 
we obtain for (\ref{eq-OO-E-basis-finite})
\begin{equation}\label{eq-OO-finite-c-expansion-coefficient-interaction}
    \langle n| {\cal OO} | n\rangle = \langle n| {\cal O} | n\rangle^2 + o\left(f(\{Q_i\})\left(\frac{c_\alpha}{y_T}\right)^{2\Delta} \right)\,.
\end{equation}
where $f(\{Q_i\}$ depends on the detailed content of the dual seed CFT (e.g. in the D1-D5 theory, it includes $Q_1$ and $Q_5$ dependence).

In \cite{Hartman:2018tkw}, at the large-N level (meaning large-$c$ in 3 bulk dimensions), the operator $(\mathcal{OO})_{c=\infty}$ was defined by subtracting self-contractions, retaining the leading term in the OPE of $\mathcal{O}(x)\mathcal{O}(y)$ as $x\to y$ that is not the identity and that survives in the strict $c=\infty$ theory.  Our goal is to approximate this in a finite-$c$ theory, including the subtraction of the would-be divergent contribution to \eqref{eq-OO-finite-c-expansion-coefficient-interaction}.   
In our explicitly regulated system, the Hilbert space is of course different from the large-$c$ one in \cite{Hartman:2018tkw}.  Below in \S\ref{sec:boundary-prescription}, we will spell out the prescription to match correlators in the regime where $GR + EFT + \dots$ applies.  
We can define the operator $\langle n|\mathcal{OO}|m\rangle$ by hand to match the correlators of $GR + EFT + \dots$ to very good approximation at large but finite $c$.  
This is a brute force method, but the essential point is that it is well-defined thanks to the power of the $T\bar T$ deformation to serve as an explicit regulator by cutting off the real spectrum in a calculable way \cite{Smirnov:2016lqw, Cavaglia:2016oda}.

Altogether, the operator \eqref{eq-OO-E-basis-finite} and, given appropriate subtractions, the deformation \eqref{eq:deformation-schematic-draft}, are well defined in the full theory.  This will be constructed by hand to complete the duality and microstate count of \cite{Coleman:2021nor}.  The full theory with mattter is not as readily exactly solvable at finite $c$ as in the case of $T\bar T+\Lambda_2$.  
This befits the system of interest, with sufficiently rich dynamics to realize large-radius gravity probed by light quantum fields.  
Still, below we will find methods to calculate physical quantities of interest, finding agreement on the two sides of the duality.

\subsection{Implication for operator algebra}

Since we have a finite system, its Von Neumann operator algebra is type $I$.  
In a strict $G_N\to 0$ limit, \cite{Chandrasekaran:2022cip}
found a type $II_1$ algebra.  Our system has finite $c$, hence finite $G_N$, and the algebra is simply type I.  

\section{Uplift and energy continuity}\label{sec:uplift}

The transition from $T\bar T+({\cal OO})_r$ to $T\bar T + ({\cal OO})_r + \Lambda_2$ involves a particular form of bulk matter.
Specifically, we should incorporate a
scalar potential that contains holographic AdS and dS minima and take into account the uplift from one to the other during the trajectory.  String-theoretic models uplifting AdS/CFT to dS include \cite{Dong:2010pm, DeLuca:2021pej}.  
The relevant matter sector for this sector in the gravity-side description is an uplift scalar $\Phi_u$ whose potential $V(\Phi_u)$ contains both an AdS minimum at $\Phi_{uA}$ and a dS minimum at $\Phi_{uS}$. In addition, to fully describe the uplift, we need to keep track of the other fields $\Phi_{\perp,i}$ that couple to the uplift scalar $\Phi_u$ and that become light in the dS phase. These fields can include heavy (even stringy) fields: as a concrete example, in the 3d string theory uplift scenario \cite{Dong:2010pm}, the uplift involves wrapping various branes on the internal space. At large internal volume compared to the 10d string scale, these stringy degrees of freedom remain below the 3d Planck scale.\footnote{The full effective potential is suppressed compared to the 3d Planck scale by powers of the string coupling and inverse volumes, as is seen by expressing the effective action in Einstein frame \cite{Silverstein:2004id}.}

For simplicity, we consider an uplift which leads to a de Sitter vacuum energy which is of the same magnitude but opposite sign to the AdS one.  It is straightforward to generalize the formulas to account for more generic dS solutions.\footnote{This will be especially motivated in generalizations of this approach to four bulk dimensions, which exhibit fine tuneability of Hubble \cite{DeLuca:2021pej}.}
The uplift step occurs at fixed $y=y_0 = 8\ell G_N/L_{c0}^2= 3/c\pi^2$.  We continuously adjust the Dirichlet boundary condition for the scalar, which is a coupling constant in the boundary theory, from $\Phi_{uA}$ to $\Phi_{uS}$.  
In addition, the low energy theory contains couplings $m_\perp(\Phi_u)^2 \Phi_\perp^2$ encoding the $\Phi_u$-dependent masses for the fields; while we gradually change $\Phi_u$ from $\Phi_{uA}$ to $\Phi_{uS}$,  $m(\Phi_u) $ decreases and the fields $\Phi_\perp$ become light. Conversely, there are many light modes in AdS that get heavy in the dS phase; similar comments apply to these.

In the gravity-side description, the classical solution space for matter excitations continuously deforms according to this continuous deformation of the scalar boundary condition (which resolves the jump in $\Lambda_2$ in \cite{Gorbenko:2018oov, Coleman:2021nor}).  After the uplift step, we proceed with the $T\bar T+({\cal OO})_r + \Lambda_2$ deformation, increasing $y$ from $y_0$ to capture the bounded cosmic horizon patch.  

The detailed uplift was not required for the pure gravity results in \cite{Coleman:2021nor}, but plays a significant role in the presence of matter, leading to continuity of the energy spectrum. 
We can see this by reviewing the behavior of the various classes of states as we evolve along the $T\bar T+\Lambda_2$ trajectory, for which $\Lambda_2$ jumps to a nonzero vallue at $y=y_0$.
 Without the $({\cal OO})_r$ contribution, the trajectory in \cite{Coleman:2021nor} does exhibit continuity of the dressed energy along the trajectory for the dressing of the seed states in the energy band $\Delta \sim c/6$ (up to very fine discontinuities that are invisible to the course-grained gravitational description).  In the gravity description, this arises from bringing in the boundary with $T\bar T$ to the horizon of a BTZ black hole, then bringing it out with $T\bar T+\Lambda_2$ to obtain the $dS_3$ cosmic horizon patch (the static patch, with boundary).  
At the junction between the two trajectories, at a dimensionless deformation parameter $y_0=\lambda_0/L^2=3/c\pi^2$ the square root term vanishes in the dressed energy
\begin{equation}\label{eq-pure-gr-energy}
  {\cal E} = \frac{1}{\pi y}\left(1\mp \sqrt{\eta +\frac{y}{y_0}(1-\eta)-4\pi^2 y (\Delta-c/12)}\right)  ~~~~ (\text{pure}~ T\bar T+\Lambda_2),
\end{equation}
so that the $\Lambda_2$-dependence (in $\eta$) drops out for the entropically dominant $\Delta \simeq c/6$ band of energies.  For that band of energies, to capture the cosmic horizon patch the branch of the square root is positive in the $T\bar T+\Lambda_2$ phase.  

 The other universal state is the vacuum, which is  captured by the original (-) sign of the square root.  In particular, as discussed in \S\ref{sec-grav-sector} we stress that  once we turn on $\Lambda_2$, our spectrum includes both signs of the square root.  In gravity-side language, this reflects the fact that the cylinder size $L_c$ is fixed, but the extrinsic curvature is not.  It can take both signs; the spectrum includes both universal sectors $\Delta \simeq c/6$ dual classically to the cosmic horizon patch, and the vacuum $\Delta =0$ dual classically to the pole patch.\footnote{The $T_t^t$ component of the Brown-York stress tensor is given by $T_t^t = \frac{1}{8\pi G_N}\left( K_t^t - ( K - 1/\ell ) \right) = \frac{1}{8\pi G_N} \left( \frac{1}{\ell} - K_\theta^\theta \right)$. For a metric of the form $ds^2 = dw^2 - g(w,t)^2 dt^2 + r(w,t)^2 d\theta^2$, $K_{\theta}^{\theta} = \frac{1}{r} \frac{\partial r}{\partial w}$ determines the square root part of $T_t^t$.}  From the boundary theory perspective, our Hamiltonian $\sum_n E_n |n\rangle \langle n |$ can include both types of levels. In the path integral formalism, a model that illustrates this very simply is the relativistic point particle. Its action $S= -m \int d\tau \sqrt{- \dot X^\mu \dot X_\mu}$ can be rewritten as a quadratic action in $\dot X_\mu$ by introducing a dynamical world-line vielbein \cite{Polchinski:1998rq}. The equation of motion for the vielbein admits two solutions corresponding to both branches of a square root.  This structure arises in the complete path integral representations of $T \bar T + \Lambda_2$, both using JT gravity \cite{Dubovsky:2018bmo} and massive gravity \cite{Tolley:2019nmm, Mazenc:2019cfg, Torroba:2022jrk}.  We note that before we turn on $\Lambda_2$, there is not a role for both signs of the square root.  In gravity language, this is because although the extrinsic curvature is not fixed, the $+$ sign corresponds to filling the patch toward the asymptotic boundary, which is not a normalizable excitation.\footnote{See however interesting works such as \cite{Griguolo:2021wgy} \cite{Iliesiu:2020zld} which consider both signs in the AdS theory.
}  

More general states with $0 < \Delta \ll  c/6$ on the other hand are not captured by the pure gravity sector.  They are model dependent; in gravity language, they can include BTZ black holes below the Hawking-Page level, particle states, and quantum field excitations.  In the gravity description, we deform from $y=y_T\ll 1$ to $y_0=8\ell G_N/L_{c0}^2=3/c\pi^2$ by changing the boundary condition governing the bulk solutions, specifically reducing the bounding proper cylinder size $L_c$ to $L_{c0}$. 
At this value $L_{c0}$, the low-lying states do not typically reach the boundary.  (If they do so at one time, they bounce off of it back into the bulk, given the ${\cal OO}$ contributions ensuring a local boundary condition.)
So (looking at the state at a generic time) the deformation to the matching point $y_0$ only brings the boundary in to a radial position outside of the black hole horizon or particle position.  If we then simply add $\Lambda_2$, without an uplift sector, and continue with the trajectory (decreasing $L_{c0}$ with $\eta=-1$ and the opposite branch of the square root) we get a discontinuity.  That is, in the absence of an explicit uplift sector, the square root term in the energy \eqref{eq-pure-gr-energy} does not vanish at $y=y_0$ in this case, so reversing its sign implies a discontinuity.  As stressed in \cite{Coleman:2021nor}, this is as expected since these low-energy model-dependent states involve the matter contributions, which should therefore be included to restore continuity.  It is this point we will establish in this section.

After adding matter contributions and the uplift sector, as discussed above there is a continuous trajectory that combines both branches. This means that both signs are contributing to the path integral. Of course, the nonlinear nature of the equations complicates the solutions matching the boundary value of $\Phi_{uc}$ in between its AdS and dS values. 
An interesting question is the degree of quantum mechanical mixing among semiclassical states with different signs of the extrinsic curvature in the full theory including matter.
A quantum mechanical model of such mixing can be given in terms of a potential with two minima, corresponding to the two branches. Small fluctuations around each minimum give rise to dressed energies $E_n^\pm$ and eigenstates $|\Psi_n^\pm \rangle$. Furthermore, the quantum-mechanical Hamiltonian contains off-diagonal terms that mix both sectors, $\langle \Psi_n^+|H|\Psi_m^- \rangle \neq 0$; these reflect the finite-energy trajectories that connect both types of states.

In addition to the states described above, our theory contains configurations which access the landscape of AdS and dS vacua given the uplift sector potential.  These are also continuous along our trajectory.  
We can construct a domain wall at a fixed time $t$ which interpolates between $\Phi_A$ and $\Phi_S$.  
For these states as well, we would like to see how the deformation including matter, as expressed for example by the trace flow equation (\ref{eq:TFE-two-vacua}), automathically yields continuity of the states making up the model-dependent sparse light spectrum.  In the gravity description, given the uplift potential we can construct a state which at a moment of time symmetry contains a  domain wall interpolating between the locally AdS and dS regions. In these variables, there is a manifestly continuous trajectory obtained by moving the Dirichlet condition on $\Phi$ to interpolate between these two values.
The boundary cylinder grows in the AdS region as we move away from the particle or BTZ horizon, and it shrinks once we continue into the dS region.  Therefore it will have a maximum somewhere at which $dg_{\theta\theta}/dw=0$ as it flips sign. This implies that the extrinsic curvature component $K^\theta_\theta$, which determines the square root part of the dressed (=Brown-York) energy formula vanishes, and it connects continuously to the opposite sign of the square root.  We stress again that in these configurations we work with Dirichlet boundary conditions corresponding to fixed intrinsic geometry. Except for simple states, this is not the same as moving the boundary in a pre-existent geometry.

\section{Summary of Boundary Theory Prescription}\label{sec:boundary-prescription}

We now express the construction of our deformed theory in $2d$ boundary variables and explain how to apply it to derive properties of the emergent geometry.  We will focus on new features and issues arising from the addition of matter.  In particular we will use the matter at the leading nontrivial order to validate and probe the emergent large-radius de Sitter geometry. 

The deformation of our theory was written in \eqref{floweq} in Lagrangian language at large $c$.  In order to make contact with the finiteness of the real spectrum at finite $c$, it is useful to recast this in Hamiltonian language
where the path integral for a transition amplitude with sources takes the form
\begin{equation}
    \langle \chi(\theta, t)|\chi'(\theta', t')\rangle_{J, K} = \int_{b.c.} D\chi D\pi_\chi \exp(i\int d^2 x [\dot\chi \pi_\chi -\mathcal{H}(\chi, \pi_\chi)+J_\chi \chi + K_\pi \pi_\chi +J \mathcal{O}])
\end{equation}
Here we use $\chi$ to denote the deformed-CFT fields.

We will include an uplift sector denoted with a subscript $u$ here, and a marginal scalar $\mathcal{O}$ with $\Delta =2$ dual to a gauge field which will realize the required approximate bulk locality.  Including the initial stint of pure $T\bar T$ to instate a finite real spectrum as explained in \S\ref{sec-finiteness}, and defining $\mathcal{S}_H = \int d^2 x \mathcal{H}$ where $\mathcal{H}$ is the Hamiltonian density of the theory,   we have 
\begin{equation}\label{eq-early-pure-TTbar}
\lambda\frac{\partial\mathcal{W}}{\partial \lambda} =  -\lambda\frac{\partial{\cal{S}_H}}{\partial \lambda} = -\int d^2 x \sqrt{-\gamma}\,2\pi\,\lambda T \bar T, ~~~~~ y<y_T
\end{equation}
\be
\begin{aligned}\label{eq-floweq-again}
\lambda\frac{\partial \mathcal{W}}{\partial \lambda} & =-\lambda\frac{\partial{\cal{S}_H}}{\partial \lambda} =\int d^2 x \sqrt{-\gamma} \frac{-1}{B_{CT}(\Phi_{uc}+J_u)} \left[2\pi \lambda T\bar{T} + \frac{1}{4}\left(\frac{\lambda c}{12}\right)^{\Delta-1}\left[(\mathcal{OO})_r+(\mathcal{O}_u \mathcal{O}_u)_r\right]
  -\frac{1}{4}\left(\frac{\lambda c}{12}\right)^{2-\Delta}(\partial_i J)^2 \right.\\
& \left. - \left(\frac{\lambda c}{48}\right)^{1/2}\left(\frac{L_c}{L}\right)^3 V(\Phi_{uc} + J_u) + \frac{1}{4\pi^2 \lambda^2} \left(\frac{\lambda c}{12}\right)^{\Delta-1} B_{CT}'(\Phi_{uc}+J_u) - \frac{B_{CT}(\Phi_{uc}+J_u)^2}{2\pi \lambda}\right]\,, \,y>y_T\,.
\end{aligned}
\ee
where we indicated renormalized operators we will shortly define, and $\Delta$ refers to the CFT dimension of the appropriate (uplift or probe) operator.
For a given seed theory, the light spectrum may contain multiple such sectors, which enter similarly in the definition of the deformation. 

For the first segment \eqref{eq-early-pure-TTbar} of our deformation, for a given total spatial momentum $p_\theta$, the Hamiltonian manifestly goes complex outside a finite region of phase space.  This is related to the energies going complex as reviewed above in \S 2.2.  It pertains to the dressed classical Hamiltonian as was computed for a free scalar seed theory:  see e.g. equations (2.16-17) of \cite{Kraus:2018xrn}.  If we restrict the path integral to this region, we respect unitarity.  Combining this with the discretenesss of the spectrum, we realize the finite real spectrum of the theory, while retaining use of the tool of continuous path integration. This formalism will be useful for deriving the operators we add along the trajectory, including the stress-energy tensor.

Once we start adding the $(\mathcal{OO})_r$ contributions, we will not be able to analytically resum the Hamiltonian in the same way as was done for the solvable $T\bar T + \Lambda_2$ deformation.  This complication is as expected: in general energy levels do not evolve independently in the more general 
case.  

However, such a resummation is not needed to define the theory.   
As in the familiar $T\bar T$ deformation we can think of this as an algorithmic step by step procedure, advancing along the trajectory in the space of theories according to \eqref{eq-early-pure-TTbar}-\eqref{eq-floweq-again} as $\lambda$ infinitesimally increases by $\Delta\lambda$.  In order to determine the operators associated to conserved currents, we may go between the Hamiltonian and Lagrangian formalisms, using the latter to derive the currents and the former to manifest the finiteness of the real energy spectrum.  In particular,
at each step in the deformation, we recompute
$T_{\mu\nu}=-\frac{2}{\sqrt{-g}} \frac{\delta}{\delta g^{\mu\nu}} S_{dressed}$ in order to add $\Delta\lambda T\bar T$ to the Hamiltonian density.  Similarly, for a sector consisting of a conserved current $\mathcal{J}$ corresponding to a global symmetry $\chi\to \chi+\Delta \chi$, $\mathcal{L}_{dressed}\to \mathcal{L}_{dressed}+\partial_\mu V^\mu$, we recompute $\mathcal{J}^\mu = \frac{\partial \mathcal{L}_{dressed}}{\partial(\partial_\mu\chi)\Delta\chi}-V^\mu$.  (We could also use the Noether procedure for the stress-energy tensor itself.)  
These fields associated with symmetries are naturally light on the gravity side and hence well motivated as appropriate probes of the bulk locality.     

In the more general case, the operators $\mathcal{O}$ can be obtained in principle as follows.  They satisfy $\langle \mathcal{O}_{I_1}\dots \mathcal{O}_{I_n}\rangle=\frac{\delta}{\delta J^{I_1}}\dots \frac{\delta}{\delta J^{I_n}}\mathcal{W}=-\frac{\delta}{\delta J^{I_1}}\dots \frac{\delta}{\delta J^{I_n}}(\mathcal{S}_H-i\int d^2 x J \mathcal{O})$. With the initial pure $T\bar T$ trajectory in place \eqref{eq-early-pure-TTbar}, $\langle \mathcal{O}(x_1)\mathcal{O}(x_2) \prod_k \mathcal{O}_k\rangle$ will be finite as we take $x_1\to x_2$.   This follows from the discreteness of the spectrum and the finite range of real energies in the phase space region of real $\mathcal{H}$.  In the finite system, the labeling of operators $\mathcal{O}(x)$ in terms of $x$ will yield redundant operators.
Thus the operator $(\mathcal{OO})_r$ that we deform by is well defined and in principle determined by calculating all of its correlators (or enough for the desired precision) at a given step in the trajectory, comparing it to the results of $GR + EFT + \dots$ when available, and explicitly subtracting deviations from the latter in order to determine what to add at the next step. 

More specifically, we can arrange this agreement as follows, taking into account the difference in Hilbert spaces between our finite theory and the formal large-N theory \cite{Hartman:2018tkw}.  Start from an energy level or band of energies visible in the universal, pure gravity, sector.  In the bounded AdS part of the trajectory, this includes states whose gravity description matches the external geometry of the BTZ black holes with masses corresponding to operator dimensions which exist in the seed CFT spectrum, along with the vacuum state matching empty AdS. In the bounded dS phase it includes states matching the pole patch and the cosmic horizon patch. From there we can construct, in our finite theory, correlators  
\begin{equation}
    \langle universal | (\mathcal{OO}(x))_r \mathcal{O}(y_1)\dots\mathcal{O}(y_m) | universal \rangle 
\end{equation}
which probe matter excitations.  These are limited in $GR + EFT + \dots$ in several ways.  The distance between points on the boundary cannot fall beneath the Planck length, $|y_i-y_j|> G_N, |x-y_i|>G_N$. The lifetime of the theory is uncertain in semiclassical $GR + EFT + \dots$ after a timescale $T\sim G_N \exp(S)$ (for entropy $S$), corresponding to energy resolution bounded by $\Delta E >G_N^{-1}\exp(-S)$.   These limitations reduce the number of independent time-ordered correlators to match to be less than order $P(S) \exp(S)$ where $P(S)$ is a power law in $S$ depending on the number of independent operators we include and the combinatorics of their insertions in the correlator.  In fact this is a conservative over-estimate of the count of correlators to match between our theory and GR:  $GR + EFT + \dots$ is only a good approximation down to length scales $\gg G_N$.  Moreover, matter containing too much entropy will condense into a black hole, thus joining the universal sector of states corresponding to pure gravity.  Altogether, we have the freedom to tune the $e^{2 S}$ matrix elements $\langle n| (\mathcal{OO})_r | m\rangle $ such that our correlators match those of $GR + EFT + \dots$ when it is under control.

Let us pause here to elaborate on the role of this tuning and how it fits into the goal of de Sitter holography.  The method in this paper is aimed at showing that a boundary theory exists which captures (i) the refined Gibbons-Hawking entropy and the radial geometry as probed by a timelike boundary, while also (ii) reproducing bulk matter dynamics.  The former (i) by itself arises in a simple way from the solvable $T\bar T+\Lambda_2$ theory.  The latter (ii) we obtain via tuning to match the set of controlled $GR + EFT + \dots$ observables, in a way that does not ruin the $T\bar T+\Lambda_2$  results (i) for entropically dominant bands of energy levels.
As discussed in \S\ref{sec:uplift} the set of degrees of freedom we keep track of extends to some internal string theoretic quantities.

The reason that tuning to match $GR + EFT + \dots$ is possible is because the number of controlled $GR + EFT + \dots$ calculations is parametrically smaller than the tuning capacity of our finite theory.  With Hilbert space dimension $\exp(S)$, the finite quantum theory has $\exp(2 S)$ matrix elements available to tune for each $GR + EFT + \dots$ operator  $\sum_{m,n} |m\rangle c_{mn}\langle n|$.  
This is true even though $GR + EFT + \dots$ naively has an infinite spectrum from a continuum of states (related to its correspondence with a strict large-N limit).  But at finite $c$ (i.e. finite 3d Newton constant $G_3$), this continuum of states is not under control in $GR + EFT + \dots$.
(As one quick check of this -- if the continuum were under control in $GR + EFT + \dots$, then AdS/CFT on $S^2\times R$, which has a discrete spectrum, would be wrong.) For example, $GR + EFT + \dots$ cannot reliably resolve black hole area differences below the Planck scale $G_3$, corresponding to mass differences of order $1/G_3$.  Similarly, the number of states accounted for by perturbative $GR + EFT + \dots$ excitations within an area $A$ is parametrically smaller than $\exp(A/4G_3)$. 

Next let us spell out the reason this will not interfere with the calculation of the refined Gibbons-Hawking entropy and radial geometry derived from the $T\bar T+\Lambda_2$ part of the deformation.  Firstly, the total state count during the last leg of the 4-step deformation (cf figure \ref{fig:trajectory-4-part-schematic}) is fixed at $\exp(S_{dS})$.  That is, one can match all of $GR + EFT + \dots$ with this state count -- the tuning part of the algorithm does not force us into a different Hilbert space dimension.  Related to this, in section 6 we will find that time-dependent matter excitations classically do not change the reality properties of the states, as their effect on the Brown-York energy multiplies the square root part of the energy.  There are also quantum contributions to the generalized entropy which depend on the matter, as would be computed in a timelike-boundary generalization of \cite{Anninos:2020hfj}.  These affect the count of real states in our theory accordingly, in a way that is subleading to the leading area term which dominates our count. Similarly, the small effects of matter of proper energy $E_{matter}$ (suppressed by $G_3 E_{matter}$) on the precise energy eigenvalues in the dominant entropy band do not change its leading behavior, so it still matches the radial geometry via the dependence of the Brown York energy on the boundary cylinder's proper size.      

We should also note here that energy levels can cross, in contrast to the situation in the integrable $T\bar T+\Lambda_2$ by itself. The Hamiltonian at a given step in the deformation can be expressed in an energy basis of our finite theory as $H=\sum_{n, j_n} | E_{n, j_n}\rangle E_n \langle E_{n, j_n} |$ with $j_n$ indexing degenerate states with energy $E_n$. At the step prior to the level crossing, this matrix has a block of the form $diag(E_{-1}, E_{-1}+\delta E)$ which becomes at the crossing point $diag(E_{0}, E_{0})$ and then splits again into $diag(E_{+}, E_{+1}+\delta E)$.  

To summarize the big picture here:  our overall proposal is to heavily exploit the fact that one does not need the holographic boundary theory to compute for us $GR + EFT + \dots$ quantities.  We do need it to explain the horizon entropy.  Our approach, in other words, highly prioritizes the calculation of the entropy via the $T\bar T+\Lambda_2$ part of the deformation, while being content with tuning to $GR + EFT + \dots$ at each step in the algorithm, to match the residual (matter) effects.

This leads to an equation for the deformation of the expectation value of the interaction Hamiltonian in the $nth$ energy eigenstate as in \cite{Smirnov:2016lqw} as 
\begin{equation}\label{eq-def-Ham}
    \partial_\lambda \langle n| H |n \rangle = 
     \langle n| 2\pi T \bar T + \frac{1}{4\lambda}\left(\left(\frac{\lambda c}{12}\right)^{\Delta_u -1}({\cal O}_u{\cal O}_u )_r+ \left(\frac{\lambda c}{12}\right)^{\Delta -1}(\mathcal{O}\mathcal{O})_r\right) |n \rangle  - \tilde{V}(\Phi_{uc}(\lambda)) - \frac{B_{CT}(\Phi_{uc}(\lambda))^2}{2\pi \lambda^2}
\end{equation}
where again the last two terms account for the uplift explained in the previous section.\footnote{In \eqref{eq-def-Ham}, $\tilde{V}$ is the term proportional to the uplift potential in \eqref{eq-floweq-again}:  $\tilde{V}(\Phi_{uc}) = \left(\frac{\lambda c}{48}\right)^{1/2}\left(\frac{L_c}{L}\right)^3 V(\Phi_{uc}).$}

For the pure $T\bar T+\Lambda_2$ theory (with no $\mathcal{OO} $ terms and with $V$ and $B_{CT}$ constant), the dressed energies $E_n=\langle n| H | n\rangle$ can be analytically determined \cite{Smirnov:2016lqw}\cite{Coleman:2021nor}.  This proceeds from either the trace flow equation \eqref{eq-EE-to-deformation} or \eqref{eq-def-Ham} by relating the independent stress-energy components $T^0_0, T^1_1,$ and $T^0_1$ to the energy, $dE/dL$, and conserved spatial momentum respectively.  Consider the $T^\theta_\theta$ component of the dressed stress energy tensor (with $\theta$ the spatial direction around the cylinder $-dt^2 +  L^2 d\theta^2$ that the boundary theory lives on).  For a homogeneous time-independent state this is the pressure $T^\theta_\theta = -dE/dL$. It is useful to work with the dimensionless deformation parameter $y=\lambda/L^2$. With these substitutions, one obtains the differential equation
\be\label{eq:Energy-diffeq}
\pi y \mc E(y) \mc E'(y)- \mc E'(y)+ \frac{\pi}{2} \mc E(y)^2=  \frac{1-\eta }{2 \pi  y^2}+2\pi^3 J^2
\ee
in terms of the dimensionless dressed energy.   
\begin{equation}\label{eq:calE}
    {\cal E} = E L\,.
\end{equation}
For a seed CFT,
\be
{\cal E}\big|_{y=0, \eta=1}={\cal E}_{CFT}=2\pi\left(\Delta -\frac{c}{12}\right)\,.
\ee
An equivalent way to determine the dressed energies is to apply the trace flow equation 
\be
T^i_i = 4\pi \lambda T\bar{T} + \frac{1}{2}\left( \left(\frac{\lambda c}{12}\right)^{\Delta_u-1} (\mathcal{O}_u \mathcal{O}_u)_r+\left(\frac{\lambda c}{12}\right)^{\Delta-1} (\mathcal{OO})_r \right)  - \Lambda_2 + \dots
\ee
Without matter, one can substitute $T^\theta_\theta=-dE/dL$ and again obtain a differential equation for $E(L)$ \cite{Gorbenko:2018oov}.

With local bulk matter,  
we must account for states with more general time and $\theta$ dependence.  In the general case, $T^\theta_\theta(t, \theta) \neq -dE/dL$ and we have the additional $(\mathcal{OO})_r \simeq \langle \mathcal{O}(t,\theta)\rangle^2 $ terms in the defining equation for the deformation.  As stressed above, the procedure is algorithmically well defined, but it is interesting to ask if we can derive a tractable differential equation for the dressed energies (and dressed operators) in the case with approximately local bulk matter.  We do not have any reason to expect a simple expression in general, but in the next section we will find this to work out in a simple class of time-dependent states (at fixed energy) which directly probe the approximate locality of the bulk matter.  Moreover, in the analysis of the $GR + EFT$ solution space, we will discover that the matter excitations lead to an effect on the energy multiplying the square root term in the dressed energy \eqref{eq-dressed-En-review-pure-GR-sec}, retaining their branch structure dictating their reality properties.

\section{Dressed energies capturing time-dependent local bulk matter}\label{sec-dressed-Es}

In this section, we derive the leading effect of matter on simple states.  We begin by deriving its structure on the boundary side of the duality.  We then derive time-dependent solutions with matter in the classical approximation on the gravity side and comment on the correspondence.  

\subsection{$2d$ boundary side derivation of dressed energies in simple cases}\label{sec-bdry-dressed-Es}

We can obtain the leading matter contribution to the dressed energies in a simple way in some cases.  Suppose we consider a state with no bulk matter excited, and then add energy by exciting the bulk matter.  Such states include oscillating solutions for the bulk matter. 

In general, we have from the defining equation of the deformation (with $T^t_\theta=0$ for simplicity)
\begin{equation}\label{eq-def-energies-Ham}
    \partial_\lambda \langle n | H | n\rangle =   L \langle n| -\frac{\pi}{2} T^t_t T^\theta_\theta + c_{\cal O} {\cal OO}-\Lambda_2 |n \rangle
\end{equation}
where the Hamiltonian $H=-L T^{t}_t$ and $c_{\mathcal{O}}$ is a constant that may depend on the operator $\mathcal{O}$ as described in \S\ref{sec:large-c}.  We are interested in capturing the effect of matter excitations at the leading nontrivial order, denoting as $\Delta T^\mu_\nu, \Delta H$ the new contributions to these quantities generated by the matter excitation.

The trace flow equation is
\begin{equation}\label{eq-tfe-glueball-sec-simpler}
    T^t_t + T^\theta_\theta = -\pi\lambda T^t_t T^\theta_\theta -2\Lambda_2 +2 c_{\cal O}\mathcal{OO}
\end{equation}

We can now substitute \eqref{eq-tfe-glueball-sec-simpler} into \eqref{eq-def-energies-Ham}, which in general gives
\begin{equation}
    \partial_\lambda E = \frac{1}{2\lambda}(LT^t_t + L T^\theta_\theta).
\end{equation}
with time-dependence arising in $T^\theta_\theta$ in the general case.

Expanding this in the effects of the matter, 
in the case of a general equation of state $\langle n | \Delta T^\theta_\theta | n\rangle =f(\lambda)\langle n | \Delta T^t_t | n\rangle$, yields the differential equation
\begin{equation}
\partial_\lambda \Delta E = -\frac{(1+f(\lambda))}{2\lambda}\Delta E
\end{equation}
with solution
\begin{equation}
    \Delta E = \Delta E(\lambda_*)\frac{\exp\left(-\int_1^\lambda\frac{(1+f(\tilde\lambda))}{2\tilde\lambda}d\tilde\lambda\right)}{\exp\left(-\int_1^{\lambda_*}\frac{(1+f(\tilde\lambda))}{2\tilde\lambda}d\tilde\lambda\right)}
\end{equation}
Thus in this situation with the expectation value of the pressure being proportional to that of the energy, the effect of matter on the energy levels is straightforward to solve for.

In the next section, we will develop the classical matter solutions  in our bounded (A)dS patches.  For a spatially homogeneous Gaussian matter variable behaving as a harmonic oscillator of a specific frequency, the expectation value of the pressure in an energy eigenstate is the same as its time average. We will find examples of matter satisfying an equation of state of the above form, with time averaging playing the role of this expectation value.

\subsection{Bulk counterpart: matter waves and bulk gravity with fixed boundary cylinder}

We next derive the emergent classical solutions describing excited matter in the bulk.  
This will exhibit important physical effects of the fixed-metric boundary condition with cylinder size $L_c$.   
This boundary condition -- which freezes the boundary geometry leaving us with a gravity-free dual theory -- is distinct from that of global de Sitter.  
A major implication of the timelike boundary deduced in \cite{Banihashemi:2022htw, Banihashemi:2022jys}  is the presence of the standard sign in the first law of thermodynamics when written in terms of the boundary energy $E_{\text{Brown-York}}$:
\begin{equation} \label{eq-first-law}
    \delta E_{\text{Brown-York}} = \delta E_{\text{bulk matter}} + T \delta S_{\text{horizon}}\,.
\end{equation}
In this section, we will illustrate the effect of matter in explicit classical solutions. In an ongoing work we are also studying quantum effects associated to imposing boundary conditions on matter at the timelike boundary. In particular, the resulting negative Casimir energy of the matter in its vacuum state has interesting effects on the causal properties of the spacetime\footnote{We thank Edgar Shaghoulian for many helpful discussions on this topic.}. See \cite{Levine_2023} for related discussions in dS JT gravity in the absence of timelike boundaries.

Let us consider the purely AdS or dS part of our trajectory (i.e. before or after the uplift steps in our deformation trajectory), and work with the action:  
\begin{equation} 
  S = \frac{1}{16\pi G_N} \int \sqrt{- g} \left( R - 2 \frac{s}{\ell^2} \right)- \int \sqrt{- g} \left( 
  \frac{1}{2} (\nabla \phi)^2 +V(\phi)\right)+S_{\text{bry}}\,.
\end{equation}
On the gravity side, we can work at the probe order, where the bulk quantum fields propagate on the un-back-reacted bounded geometry, up to corrections suppressed by powers of $G_N$ times gradients. 
In order to capture the Brown-York stress-energy
\begin{equation}\label{eq-BYdef-in-tdep-section}
    T_{ij} =- \frac{1}{8\pi G_N} \left(K_{ij} - \left(K-\frac{1}{\ell}\right)\gamma_{ij}\right)
\end{equation}
in this probe regime,
we do need to calculate the order $G_N$ back reaction on the metric.
The system of probe scalar and back reacted metric have simple standing wave solutions satisfying fixed boundary conditions.   We stress again that in the general time-dependent setting with local bulk matter, Changing $L_c$ is not the same as moving a cylinder within a solution for the original $L_c$ to adopt as a new boundary, since the quantization of modes depends on $L_c$. In other words, we are not `moving in the boundary' in a pre-existing solution, but are instead changing its proper size.


We can organize the calculation in an expansion in $G_N$ times the energy scale of the bulk EFT matter, and will work in the leading nontrivial order in this expansion, as follows.  We treat the bulk scalar as a probe of the original geometry, but we include the back reaction on gravity necessary to capture the order 1 contributions to both the Brown-York energy and the generalized entropy. That is, we compute the order $G_N$ contribution to the extrinsic curvature $K$ at our boundary, and the order $G_N$ contribution to the horizon area ${\cal A}$.

In this section, we summarize the calculation of the metric back-reaction.\footnote{Our analysis generalizes the linearized part of the analysis of \cite{Bizon:2011gg} to the case with finite timelike boundary in (A)dS. Cf. \cite{He:2023hoj, He:2023knl} for a perturbative analysis of gravity+scalar systems in AdS$_3$ with an emphasis on the computation of torus correlators.  }  We consider for simplicity backreacted configurations that preserve a $U(1)$ symmetry around the $\theta$ circle, and we display here an explicit class of solutions on bounded patch that includes the de Sitter observer, imposing Dirichlet conditions at the boundary of this patch. We work with the metric ansatz of the form:
\begin{equation} \label{eq:ansatzMetrho}
ds^2_3 = \frac{4}{(1+s\frac{\rho^2}{\ell^2})^2}\left(-\rho^2 B e^{-2D} d t^2+B^{-1} d\rho^2+\frac{\ell^2}{4 }\left(1-s\frac{\rho^2}{\ell^2}\right)^2 d\theta^2\right)\,.
\end{equation}
When $B=1$ and $D=0$, this metric corresponds to BTZ, Rindler, and the static patch for $s=-1, 0,$ and $+1$ respectively. In these coordinates, the pole of the static patch is at $\rho=\ell$, and the horizon is at $\rho=0$. We will study perturbations around the background geometry including those sourced by matter excitations by writing (with $\kappa^2 = 16\pi G_N$)
\begin{equation}
     B = 1+\kappa^2 \Delta B(t,\rho)\,,\qquad D = 0+\kappa^2\Delta D(t,\rho)\,.
\end{equation}

Before we proceed to the calculations of the back-reactions $\Delta B$ and $\Delta D$, we note that the Brown-York energy of the pole patch is given as
\begin{equation} \label{eq-BY-bulk-rho}
     T_t^t=\frac{-1}{8\pi G_N \ell} \left( 1 \pm \frac{2 s \rho_c \ell}{\ell^2 - s\rho_c^2}\sqrt{1+\kappa^2 \Delta B}\right)\,
\end{equation}
for a boundary situated at $\rho=\rho_c$. The first factor in the second term corresponds to the square root part of the dressed energy \eqref{eq-pure-gr-energy} in the $T\bar T + \Lambda_2$ deformed theory.  We can see this as follows. The relationship between the $\rho$ coordinate in \eqref{eq:ansatzMetrho} and the radial $r$ coordinate in the usual BTZ/static patch metric, 
\be
ds^2_{\mathrm{BTZ/Static}} = -\frac{s(r_h^2 - r^2)}{\ell^2}d\tau^2 + \frac{s\ell^2}{r_h^2 - r^2} dr^2 + r^2 d\theta^2\,,
\ee
is given as
\begin{equation}
  \frac{r}{r_h}= \frac{\ell^2+\rho^2}{\ell^2-\rho^2}, 
\end{equation}
and applying the dictionary \eqref{eq:param-dictionary-matter} we have, for example for the BTZ black hole at the Hawking-Page level ($r_h = \ell$ and $\Delta = c/6$),
\begin{equation}
    \frac{2 \rho_c \ell}{\ell^2 - s\rho_c^2} \, \Leftrightarrow \, \sqrt{1 - \frac{4\pi^2\ell^2}{L_c^2}} =\sqrt{1-4\pi^2 \frac{8G_N\ell}{L_c^2}\frac{\ell}{8G_N}}\,\Leftrightarrow\, \sqrt{1-4\pi^2 y\left(\frac{c}{12}\right)}\, .
\end{equation}
The last expression is the square root in the dressed energy formula for $\Delta =c/6$.
Therefore, from \eqref{eq-BY-bulk-rho}, we see that the effect of the matter multiplies the original square root and does not affect the reality of the spectrum at the level of the classical solutions, as we advertised earlier.

In the following, we will solve the linearized equation in specific situations. Before doing that, we discuss here their general structure.
Upon linearization, we have $1$ equation for the scalar ($E_\phi$)  and we organize the Einstein equations as the Hamiltonian constraint ($\mathcal{H}_0 \equiv E_{0,0}$), the $2$ momentum constraints ($\mathcal{H}_i \equiv E_{0,i}$), one which is automatically zero in our ansatz because of translation symmetry along the circle, and the remaining Einstein equations $E_{\rho, \rho}$ and $E_{\theta, \theta}$, so that we have 5 total equations.
The Hamiltonian constraint is preserved by time evolution. 
Indeed, combining the linearized equations one can see that there exist $c_i(\rho)$ such that
\begin{equation}
    \partial_t \mathcal{H}_0 = c_1(\rho)\partial_\rho \mathcal{H}_1+c_2(\rho) \mathcal{H}_1+c_3(\rho) E_\phi. 
\end{equation}
Thus, we only need to enforce $\mathcal{H}_0$ at the initial time.
In addition, for $s \neq 0$,  $E_{\theta,\theta}$ is redundant, being a combination of the other other ones, their derivatives, and the equation for the scalar.
Summing up, our linearized system is composed by the equation for the scalar ($E_\phi$), a constraint equation for the metric initial data ($\mathcal{H}_0$), and three other equations.

\subsubsection{Scalar solutions and bakcreactions in the Pole Patch}
\begin{figure}[ht]
    \centering
    \begin{subfigure}[b]{0.55\textwidth}
         \includegraphics[width=\textwidth]{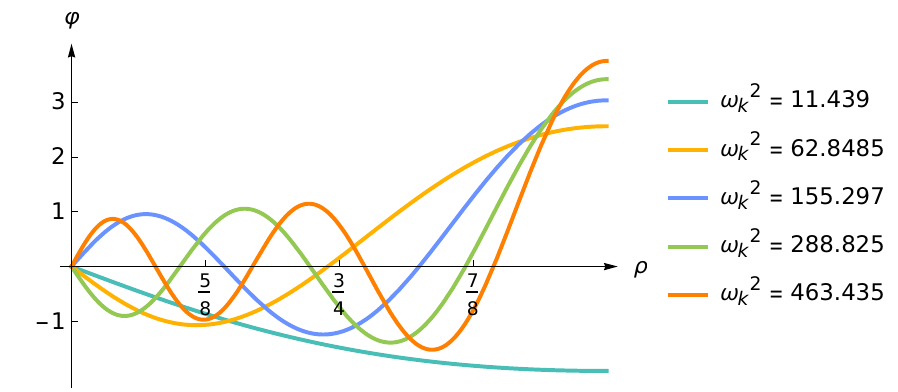}
        \caption{First 5 scalar modes for $\rho_c = \frac{1}{2}$}
        \label{fig:figure1}
    \end{subfigure}
    \hfill 
    \begin{subfigure}[b]{0.44\textwidth}
        \includegraphics[width=\textwidth]{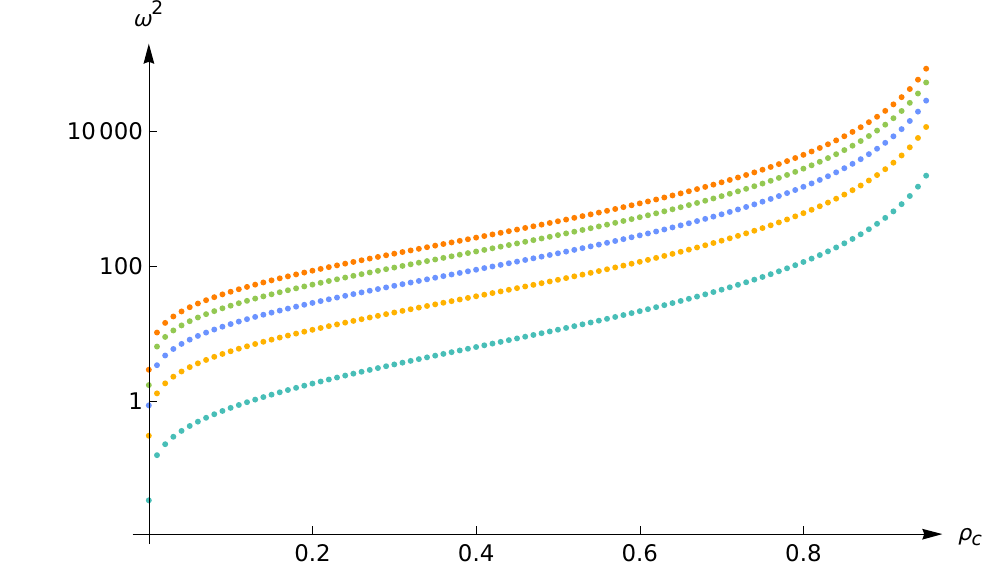}
        \caption{Behavior of the first 5 scalar frequencies as function of $\rho_c$ (which determines $L_c$ in the metric \eqref{eq:ansatzMetrho}).}
        \label{fig:figure2}
    \end{subfigure}
    \caption{Example modes solving the scalar equation \eqref{eq:SLscal} for $\ell = 1$.}
    \label{fig:scal}
\end{figure}

We start by considering fluctuations of the scalar around the background, and we choose $s=+1$ and work with the pole patch $\rho\in \left[ \rho_c ,\ell \right]$ in this section. Due to linearity of the scalar equation, we can decompose a generic scalar fluctuation as a superposition of modes
\begin{equation}
\phi_1(t,\rho) \equiv \sum_k c_k e^{i \omega_k t}\varphi_k(\rho)    
\end{equation}
and we impose regularity at the pole by requiring $\varphi'(\ell) = 0 $ and the Dirichlet condition at $\rho = \rho_c$ by setting $\varphi_(\rho_c) = 0$. With these conditions, the set of frequencies $\omega_k^2$ is discrete and dependent on the cylinder size $L_c=L(\rho_c)$.  We stress here again the difference between our prescription and `bringing in the boundary' in a pre-existing solution.  Fixing the proper size $L_c$ of the cylinder for each point along our trajectory leads to new solutions with different scalar frequencies as we evolve along the trajectory by changing $y=8\ell G_N/L_c^2$. Explicitly, the equation of motion for a minimally-coupled massless scalar reduces to the eigenvalue problem
\begin{equation}\label{eq:SLscal}
-\partial_\rho\left(\rho\frac{\rho^2-\ell^2}{\rho^2+\ell^2} \partial_\rho \varphi_k\right) = \omega^2_k \frac{\rho^2-\ell^2}{\rho(\ell^2+\rho^2)} \varphi_k
\end{equation}
with boundary conditions $\partial_\rho\varphi_k(\ell) = 0$ (regularity) and $\varphi_k(\rho_c) = 0$ (Dirichlet). Some examples of solutions are shown in Fig.~\ref{fig:scal}.

We can now turn to the equations for the metric backreaction. We have one constraint equation plus two equations for the metric perturbations $\Delta B$ and $\Delta D$. Explicitly, they read 
\begin{subequations}\label{eq:eomsBackPole}
\begin{align}
\partial_\rho  \left( \frac{\rho^2}{(\ell^2+\rho^2)^2}\Delta B \right) &=     \frac{\left(\ell^2-\rho ^2\right) \left(\rho ^2 \left(\phi
   _1'\right){}^2+\dot{\phi _1}{}^2\right)}{8 \ell^2 \rho 
   \left(\ell^2+\rho ^2\right)} = \frac{\mathcal{E}^\phi}{4\ell^3} \label{eq:eomsBackPole-B}\\
   \partial_t  \left( \frac{\rho^2}{(\ell^2+\rho^2)^2}\Delta B \right)& = \frac{\rho   \left(\ell^2-\rho ^2\right)
   \phi _1' \dot{\phi _1}}{4 \ell^2 \left(\ell^2+\rho ^2\right)}\\
  \partial_\rho \Delta D &= \frac{\left(\ell^4-\rho ^4\right) \left(\rho ^2 \left(\phi
   _1'\right){}^2+\dot{\phi _1}{}^2\right)}{8 \ell^2 \rho ^3} = \frac{(\ell^2+\rho^2)^2}{4\ell^3 \rho^2} \mathcal{E}^\phi\,.  \label{eq:eomsBackPole-D}
\end{align}    
\end{subequations}
Here $\mathcal{E}$ is related to the energy density of the scalar field,
\begin{equation}
    \mathcal{E}^\phi= \sqrt{-g} u^\mu u^\nu \mathcal{T}^\phi_{\mu \nu}\,,\qquad \mathcal{T}_{\mu \nu}^\phi \equiv -\frac{2}{\sqrt{-g}}\frac{\delta S}{\delta g^{\mu\nu}} = \nabla_\mu \phi \nabla_\nu \phi -\frac{1}{2} g_{\mu \nu}\left(2 V+\left(\nabla\phi\right)^2\right)
\end{equation}
where $u^\mu$ is a unit future-pointing time-like normal vector.

The first equation is the Hamiltonian constraint, and once imposed at $t = 0$, it is automathically solved for any $t$, provided the other equations are solved. To display an explicit backreacted solution, we now choose as initial condition for the scalar $\phi_1(0,\rho) = 0$ and we consider single mode solutions. That is, we write the scalar perturbation as
\begin{equation}
\phi_1(t,\rho) \equiv \sin(\omega_k t)\varphi_k(\rho),
\end{equation}
for a given fixed pair $(\omega_k, \varphi_k)$.

Since the appearance of the scalar in the metric equations of motion is quadratic, a single mode will give rise to a piece constant in time and a piece proportional to $\sin^2(\omega_k t)$. We can thus parametrize the backreaction as
\begin{align}
    \Delta B(t,\rho) \equiv \Delta B_0(\rho) + \sin^2(\omega_k t) \Delta B_1(\rho)\\
    \Delta D(t,\rho) \equiv \Delta D_0(\rho) + \sin^2(\omega_k t) \Delta D_1(\rho)
\end{align}
Plugging in the equations of motion \eqref{eq:eomsBackPole} we find
\begin{equation} \label{eq:Delta-B1-eq}
    \Delta B_1 = \frac{\left(\ell^4-\rho ^4\right) \varphi_k \varphi_k '}{8 \ell^2 \rho }
\end{equation}
and the system of ODEs
\begin{subequations} \label{eq:Delta-BD-ODE}
\begin{align}
   \partial_\rho\left(\frac{\rho^2}{(\ell^2+\rho^2)^2}\Delta B_0\right) & = \frac{\varphi_k ^2 \omega_k ^2 (\ell-\rho ) (\ell+\rho )}{8 \ell^2 \rho  \left(\ell^2+\rho ^2\right)}\\
    \Delta D_0 ' &= \frac{\varphi_k ^2 \omega_k ^2 \left(\ell^4-\rho ^4\right)}{8 \ell^2 \rho ^3}\\
    \Delta D_1 ' & = - \frac{\left(\ell^4-\rho ^4\right) \left(\rho ^2 \left(\varphi_k '\right)^2-\varphi_k ^2 \omega_k ^2\right)}{8 \ell^2 \rho ^3}
\end{align}    
\end{subequations}
When solving these equations, we impose regularity of the solution at the pole 
\begin{equation}
    \Delta B_0(t,\ell) = \Delta B_1(t,\ell) = 0.
\end{equation}
Finally, to enforce the Dirichlet condition at $\rho = \rho_c$, we also impose
\begin{equation} \label{eq:Dirichlet-bc}
  \left. \Delta g_{tt} \right|_{\rho=\rho_c} =0 \quad \Leftrightarrow \quad \Delta B_0 (t,\rho_c)= 2\Delta D_0 (t,\rho_c),\, \Delta B_1 (t,\rho_c)= 2\Delta D_1 (t,\rho_c)
\end{equation}

\begin{figure}[ht]
    \centering
\includegraphics[width=\textwidth]{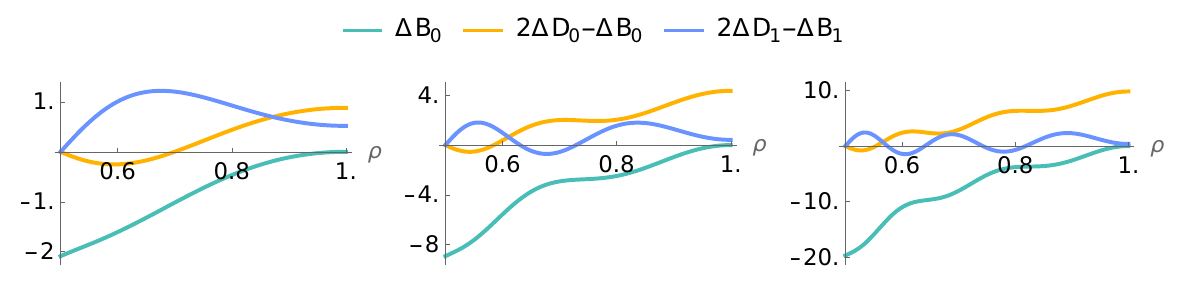}
        \caption{Metric backreaction with Dirichlet boundary conditions for the first three scalar modes ($\rho_c = 0.5 , \ell = 1$).}
        \label{fig:deltaBD-Dirichlet}
\end{figure}

Given \eqref{eq:Delta-B1-eq}, the regularity of $\Delta B_1$ at the pole is automatically satisfied if the scalar field is regular. There are three integration constants to fix when solving the ODE's \eqref{eq:Delta-BD-ODE}. The regularity of $\Delta B_0$ at the pole fixes one. Imposing \eqref{eq:Dirichlet-bc} fixes the remaining two integration constants.
After imposing the regularity condition and the Dirichlet condition for the time-dependent part of $\Delta g_{tt}$ (setting $\Delta B_1 = 2\Delta D_1$ at $\rho=\rho_c$), we get
\begin{subequations}\label{eq:solBackModes}
\begin{align}
        \Delta B_1 &= \frac{\left(\ell^4 - \rho^4 \right) \varphi(\rho)  \varphi '(\rho)}{8 \ell^2 \rho}\\
    \Delta B_0 &=     \frac{(\ell^2+\rho^2)^2 }{\rho^2}\int_{\ell}^\rho \frac{\omega_k ^2 \left(\ell^2-\tilde{\rho}^2\right) \varphi (\tilde{\rho})^2}{8 \ell^2 \tilde{\rho}  \left(\ell^2+\tilde{\rho}^2\right)}d\tilde{\rho}\\
    \label{eq:D1}
    \Delta D_1 &= \int_{\rho_c}^\rho \frac{(\ell^2-\tilde{\rho}^2 )^2 \varphi (\tilde\rho ) \varphi '(\tilde\rho )}{4 \ell^2 \tilde{\rho}^2} d\tilde{\rho} +\frac{\left(\ell^4-\rho^4\right) \varphi (\rho ) \varphi '(\rho )}{8 \ell^2 \rho }\\
    \label{eq:D0}
    \Delta D_0 &= \int_{\rho_c}^\rho\frac{\omega_k ^2 \varphi(\tilde{\rho}) ^2 \left(\ell^4-\tilde{\rho}^4\right)}{8 \ell^2 \tilde{\rho}^3} d\tilde\rho + \gamma_0
\end{align}
\end{subequations}
as the solutions to the equations \eqref{eq:Delta-BD-ODE}. 
Fully imposing the Dirichlet condition \eqref{eq:Dirichlet-bc} then fixes the remaining integration constant as
\begin{equation}
    \gamma_0 = \frac{(\ell^2 + \rho_c^2)^2}{2\rho_c^2} \int_\ell^{\rho_c} \frac{\omega_k^2(\ell^2-\tilde{\rho}^2)\varphi(\tilde{\rho})^2}{8\ell^2 \tilde{\rho}(\ell^2 + \tilde{\rho}^2)} d\tilde{\rho}\,,
\end{equation}
completely determining the solutions. Fig.~\ref{fig:deltaBD-Dirichlet} shows some plots of the metric back-reaction for the first three modes in this case.

We can compare the time-averaged change in Brown-York pressure and energy on these back-reacted solutions, in order to obtain an equation of state.
Explicitly, the change in the $(t,t)$ and $(\theta,\theta)$ components of the BY tensor after taking backreaction into account is given as
\begin{align}
    \Delta T_t^t &= \frac{-2\rho_c \Delta B(\rho_c)}{\ell^2-\rho_c^2}\\
    \Delta T_{\theta}^{\theta} &= \frac{1}{2\ell^2 \rho_c}\left((\ell^2 - \rho_c^2)\Delta B(\rho_c) + \rho_c (\ell^2 + \rho_c^2)(\Delta B'(\rho_c) - 2\Delta D'(\rho_c))\right)\,.
\end{align}
Using the Einstein's equations \eqref{eq:eomsBackPole-B} and \eqref{eq:eomsBackPole-D}, we have
\begin{equation}
    \rho_c (\ell^2+\rho_c^2) \left(\Delta B'(\rho_c) - 2\Delta D'(\rho_c)\right) = -2( \ell^2-\rho_c^2)\Delta B(\rho_c) - \frac{(\ell^2-\rho_c^2)(\ell^2 + \rho_c^2)^2}{16\ell^4 \rho_c} \phi^{\prime}(\rho_c)^2\,.
\end{equation}
Combining the results, we can relate the order-1 change in the BY pressure and the energy density as
\begin{equation}
    \Delta T^{\theta}_{\theta} =  \frac{(\ell^2 - \rho_c^2)^2}{4\ell^2 \rho_c^2}  \Delta T^t_t - \frac{(\ell^2-\rho_c^2)(\ell^2+ \rho_c^2)^2}{16\ell^4\rho_c} \phi' (\rho_c)^2\,.
\end{equation}
We consider a single-mode scalar with $\phi = \sin(\omega t)\varphi(\rho)$ and time-average the above equation over a period ($\int_0^{2\pi/\omega} dt$) to get
\begin{equation}\label{eq:quasi-state}
    \langle \Delta T^{\theta}_{\theta} \rangle =  \frac{(\ell^2 - \rho_c^2)^2}{4\ell^2 \rho_c^2} \langle \Delta T^t_t \rangle - \frac{\pi(\ell^2-\rho_c^2)(\ell^2+ \rho_c^2)^2}{16\ell^4\rho_c\omega_k} \varphi' (\rho_c)^2
\end{equation}
The terms $\langle \Delta T^{\theta}_{\theta} \rangle$ and $\langle\Delta T^{t}_{t}\rangle$ are proportional to $\omega$, as a consequence of the relevant metric backreaction being proportional to $\omega^2$, as seen from \eqref{eq:solBackModes}, and of the time-averaging introducing an $\omega^{-1}$ factor. For large $\omega$, the final term in \eqref{eq:quasi-state} is also proportional to $\omega$ , by virtue of the fact that $\varphi'(\rho_c) \propto \omega$. This latter statement can be understood from the large frequency behavior of the scalar modes, which for large $\omega$ approach the flat space oscillatory behavior $\sim e^{i \omega x}$, or numerically by using the explicit mode solutions of \eqref{eq:SLscal}. This assumes a common normalization of the solutions to \eqref{eq:SLscal}. Regardless of the normalization we can consider ratios, and combining the arguments above we obtain the equation of state of the form
\begin{equation}
    \frac{\langle \Delta T^{\theta}_{\theta} \rangle}{\langle\Delta T^{t}_{t}\rangle}  \xrightarrow{k \gg 1} f(\rho_c/\ell)\;,
\end{equation}
where $k$ denotes the index of the mode, with frequency $\omega_k$.  This relation realizes the case discussed above in boundary language in \S\ref{sec-bdry-dressed-Es}.

\subsection{First law of thermodynamics for the Cosmic Horizon patch}
In this section, we study the back-reaction in the cosmic horizon patch and verify that the solutions satisfy the first law of thermodynamics \eqref{eq-first-law} in \cite{Banihashemi:2022htw,Banihashemi:2022jys} at the probe level. We modify the metric ansatz a little by writing the back-reaction in terms of $f_1(t,\rho),f_2(t,\rho)$ defined as
\begin{equation}
    B(t,\rho) = f_1(t,\rho) \left( \frac{(\ell^2 + s\rho^2)^2}{4\rho^2\ell^2}\right), \quad D(t,\rho) = \frac{1}{2} \log \left( \frac{f_1 (t,\rho)}{f_2 (t,\rho)}\right)\,,
\end{equation}
in terms of which the metric ansatz reads
\begin{equation}
    ds^2 = -s\ell^2 f_2 dt^2 + \frac{16\ell^6 \rho^2}{(\ell^2 +s\rho^2)^4 f_1}d\rho^2 + \ell^2 \frac{(\ell^2 - s\rho^2)^2}{(\ell^2 + s\rho^2)^2}d\theta^2\,.
\end{equation}
On the background, we have
\begin{equation}
    f_2^{(0)} = 1- \frac{(\ell^2 - s\rho^2)^2}{(\ell^2 + s\rho^2)^2},\quad f_1^{(0)} = sf_2^{(0)}\,,
\end{equation}
and the zero-th order solution describes the static patch/BTZ for $s=+1/-1$ as before. To work with the cosmic horizon patch, we set $s=1$ and $\rho\in \left[0,\rho_c\right]$, and impose Dirichlet boundary conditions for the metric back-reaction at $\rho=\rho_c$. We look for linearized solutions of the form
\begin{equation}
    f_1 = f_1^{(0)} + \kappa^2 \Delta f_1,\quad f_2 = f_2^{(0)} + \kappa^2 \Delta f_2\,.
\end{equation}
The change in the horizon entropy $\Delta S = \Delta A/4G_N$ is captured by a shift in the position of the horizon from $\rho=0$ to $\rho=\rho_h$. To find this shift, we need a solution for $\Delta f_1$, which can be found by integrating
\begin{equation}
    \partial_\rho \Delta f_1 = \frac{(\ell^2 - \rho^2)(\rho^2\phi^{\prime 2} + \dot\phi^2)}{2\rho(\ell^2 + \rho^2)}=\frac{\mathcal{E}^\phi}{\ell}\,.
\end{equation}
Vanishing of $g^{\rho\rho}$ at the new horizon $\rho_h$ means that we have
\begin{equation} \label{eq:new-horizon-defn}
    f_1(t,\rho_h) = \frac{4\ell^2 \rho_h^2}{(\ell^2 + \rho_h^2)^2} + \kappa^2 \Delta f_1 = 0\,.
\end{equation}
The solution for $\Delta f_1$ can be written as
\begin{equation}
    \Delta f_1(t,\rho) = \int_{\rho_c}^\rho \frac{\mathcal{E}^\phi}{\ell} d\tilde\rho + C_{f_1} \,,
\end{equation}
and using this solution, \eqref{eq:new-horizon-defn} becomes
\begin{equation}\label{eq:new-horizon-defn-2}
    \frac{4\ell^2 \rho_h^2}{(\ell^2 + \rho_h^2)^2} + \kappa^2 \int_{\rho_c}^{\rho_h} \frac{\mathcal{E}^\phi}{\ell} d\rho + \kappa^2 C_{f_1} = 0\,.
\end{equation}
Each term in \eqref{eq:new-horizon-defn-2} can be re-written using
\begin{subequations}
\begin{align}
    T\Delta S_{hor} &= \frac{T\Delta \mathcal{A}}{4G_N} = -\frac{\rho_h^2}{4G_N \ell \rho_c}\frac{\ell^2+\rho_c^2}{\ell^2+\rho_h^2},\\ 
    \Delta E_{mat } &= \frac{\pi(\ell^2+\rho_c^2)}{\ell \rho_c} \int_{\rho_h}^{\rho_c} \frac{\mathcal{E}^\phi}{\ell} d\rho,\\
    \Delta E_{BY} &= \frac{\pi C_{f_1}(\ell^2 + \rho_c^2)}{\ell \rho_c }
\end{align}
\end{subequations}
to yield
\begin{equation}
    \frac{\kappa^2\ell \rho_c}{\pi(\ell^2 + \rho_c^2)} \left(-\frac{\ell^2 \,T\Delta S_{hor}}{\ell^2 + \rho_h^2} - \Delta E_{mat} + \Delta E_{BY}\right) = 0\,.
\end{equation}
The shift in the horizon $\rho_h^2\sim G_N$ is suppressed in our approximation scheme, and therefore, at the probe level, the above equation implies
\begin{equation}
    \Delta E_{BY} = \Delta E_{mat} + T\Delta S_{hor}
\end{equation}
confirming that the first law of thermodynamics \eqref{eq-first-law} holds for our solutions, with the positive sign \cite{Banihashemi:2022htw, Banihashemi:2022jys}.

\section{$T^2+J^2+\Lambda_2$ and a charged static patch horizon}

A special time-independent case of the $T\bar{T}+\mathcal{O}^2+\Lambda_2$ deformation is the case where $\mathcal{O}$ corresponds to a conserved $U(1)$ current $J$ in the boundary theory. A CFT deformed this way has been discussed previously in \cite{Hartman:2018tkw} as being dual to a $U(1)$ gauge field $A_{\mu}$ in the bulk, with $J$ being the corresponding conserved current. At large $c$ $\ev{J^2}$ factorizes, and one can solve for the deformed energies in closed form. Here we present this solution for $d=2$. For $d>2$, the deformed energies are solved for in \cite{Hartman:2018tkw}.

The deformation of the Lagrangian is given by
\begin{align}
    \frac{\partial \mathcal{L}}{\partial\lambda}=-2\pi T\bar{T} \mp\frac{\lambda_J}{\lambda} J\bar{J}+ \frac{1-\eta}{2\pi \lambda^2}.
\end{align}
The differential equation for the deformed energies is computed using the Hellman-Feynman theorem:
\begin{align}
    \frac{\partial E_n}{\partial\lambda}
    &=\bra{n}\partial_{\lambda}H\ket{n}\\
    &= L\left(2\pi\langle T\bar{T}\rangle \pm \frac{\lambda_J}{\lambda}\langle J\bar{J}\rangle - \frac{1-\eta}{2\pi\lambda^2}\right)\\
    &=-\frac{2\pi}{4}\left(E_n\frac{\partial E_n}{\partial L} +\frac{P_n^2}{L}\right) \pm \frac{Q^2}{\lambda L} -\frac{(1-\eta)L}{2\pi\lambda^2},
\end{align}
where we used that in Euclidean signature $H_{int}=\int d\phi \mathcal{L}_{int}$, and $\ev{J\bar{J}}=\ev{J}\ev{\bar{J}}=\frac{Q_1Q_2}{L^2}=\frac{Q^2}{\lambda_JL^2}$ at large $c$. In terms of the dimensionless energy $\mathcal{E}_n = E_n L$ and $y = \lambda/L^2$, the solutions to the above differential equation are
\begin{equation}
\label{eq:J2_deformed}
\mathcal{E}_n(y) = \frac{1}{\pi y}\left[1\pm \sqrt{\eta + C_1 y +4 \pi ^4 J_n^2 y^2\mp 2 \pi  Q^2 y \log (y)}\right],
\end{equation}
with $C_1$ a constant. Here, we have the freedom to choose the sign in front of the first square root, and the sign in the square root is fixed by the sign of the $J\bar{J}$ term in the flow equation. The above solution differs from the pure $T\bar{T}$ solution in the extra $y\log(y)$ term inside the square root.

These deformed energies are reproduced via a bulk analysis by considering a 3d Einstein plus Maxwell theory with cosmological constant
\begin{align}
    S=\int d^3x\left[\frac{1}{16\pi G}\left(R+2\frac{\eta}{\ell^2}-\frac{1}{4}F_{\mu\nu}F^{\mu\nu}\right)\right].
\end{align}
We consider Dirichlet boundary conditions for the gauge field $A_{\mu}$, so that the variation of the action with respect to $A$ at the boundary goes to zero
\begin{align}
    \delta_A S_{\text{bdry}}=-\int d^3x\sqrt{-g}\partial_{\mu}(\delta A_{\nu}F^{\mu\nu})=0,
\end{align}
eliminating the need to add a boundary term associated to the Maxwell field to the action. The solution to the corresponding equations of motion is the Reissner-Nordstrom black hole in AdS, and the static patch with a charged horizon in dS. The metric is \cite{Martinez_2000}
\begin{align}
    ds^2=-f(r)dt^2+\frac{dr^2}{f(r)}+r^2\left(d\phi^2-\frac{4GJ}{r^2}dt\right)^2,
\end{align}
where
\begin{align}
    f(r)=-8GM+\eta\frac{r^2}{\ell^2}+\frac{16G^2J^2}{r^2}-8\pi G q^2\log\frac{r}{\ell}.
\end{align}
In $d=2$ the electric charge renormalizes the mass, and we can write $f(r)$ in terms of the renormalized mass at some scale $r_0$ \cite{Martinez_2000},
\begin{align}
    f(r)&=-8GM(r_0)+\eta \frac{r^2}{\ell^2}+\frac{16G^2J^2}{r^2}-8\pi Gq^2\log\frac{r}{r_0},\\
    M(r_0)&=M+\pi q^2 \log\frac{r_0}{\ell}.
\end{align}
To derive the deformed energies, we again use the holographic stress tensor
\begin{align}
    T_{ij}=\frac{-1}{8\pi G}\left(K_{ij}-Kg_{ij}+\frac{1}{l}g_{ij}\right)
\end{align}
to get
\begin{equation}
    -T^t_t=\frac{1}{8\pi Gl}\left(1-\frac{\ell}{r_c}f(r_c)^{1/2}\right) = \frac{1}{8\pi G \ell}\left( 1 - \sqrt{\eta - 8G M\frac{\ell^2}{r_c^2} - 8\pi G q^2\frac{\ell^2}{r_c^2} \log \left(\frac{r_c}{r_0}\right)}\right) \,.
\end{equation}
We note that this energy matches with \eqref{eq:J2_deformed} upon applying \eqref{eq:param-dictionary-matter} and identifying $\pi^2 \ell q^2 = Q^2$ to get
\begin{equation}
    -L_c^2 T^t_t = \frac{1}{\pi y} \left( 1 - \sqrt{\eta - 2\pi y\mathcal{E}_n - 2\pi Q^2 y \log \left( \frac{y_0}{y} \right)}\right)\,
\end{equation}
where we defined $y_0 = 8G\ell/(2\pi r_0)^2$ so that
\begin{equation}
\log\left(\frac{r_c}{r_0}\right) = \frac{1}{2}\log\left(\frac{L_c^2}{L_0^2}\right) = \frac{1}{2}\log\left(\frac{y_0}{y}\right)  \,.
\end{equation}
This evaluates to a result that matches Eq. (\ref{eq:J2_deformed}):
\begin{align}
    -T^t_t =\frac{1}{\pi \lambda}\left(1-\sqrt{\eta-\ell M(r_0)\frac{\lambda}{r^2}-\pi q^2 \frac{\lambda}{r^2}\log\frac{r}{r_0}+\frac{1}{4}\frac{\lambda^2 J^2}{r^4}}\right)
\end{align}

\section{Summary and Discussion}

In this work, we have developed a concrete proposal for a holographic dual definition of weakly curved three-dimensional gravity plus matter with positive cosmological constant, in the presence of a non-gravitational timelike boundary.  This finite quantum system validates the microstate count obtained from the universal and solvable $T\bar T+\Lambda_2$ deformation in \cite{Coleman:2021nor} (matching \cite{Gibbons:1977mu, Anninos:2020hfj}) by embedding the latter into a theory which captures effective field theory matter with local boundary conditions. 
  
The algorithm we presented to construct the theory prescribes brute-force tuning steps which are possible thanks to various technical developments in the field. 
In particular, the utility \cite{Zamolodchikov:2004ce} of specifying theories through a differential equation extends beyond the simple solvable $T\bar T$ case \cite{Hartman:2018tkw}, but requires extra ingredients to address bulk matter.  We can work with an explicitly regulated, finite theory using the cutoff on the real spectrum from pure $T\bar T$ as the first, infinitesimal step in our procedure.  With this regulator in place, we can tune away divergences in composite operators required to deform bulk matter boundary conditions, matching to their large-N definition (worked out in \cite{Hartman:2018tkw} and generalized here).  This enables one to formulate the boundary theory as in algorithm \ref{algo}.  This theory matches, by construction, $GR + EFT + \dots$ where that description applies, and defines the quantum gravity theory beyond that (including at the bounding wall, which is a UV-sensitive ingredient).
Along the way, we analyzed the solution space of bulk matter interacting with gravity, illustrating their interplay and reproducing the thermodynamic relations \cite{Banihashemi:2022htw, Banihashemi:2022jys} appropriate to the dual as an ordinary quantum system:  e.g. the first law of thermodynamics takes the standard form \eqref{eq-first-law} in terms of the quasilocal (dressed) energy.    

A key outcome is, as already emphasized, the confirmation that the Gibbons-Hawking entropy admits a precise microstate count explanation \cite{Coleman:2021nor}, in a finite quantum system with a type I Von Neumann algebra which captures the local dS geometry.
Although defined concretely by algorithm \ref{algo}, we do not have a closed-form expression for the Hamiltonian.  
Still, we found some observables -- such as the effect of matter excitations with a simple equation of state on dressed energies -- whose structure is calculable on both sides. More could likely be done along those lines.  
Moreover, as already stressed, our theory contains a core, universal sector matching de Sitter gravity that is described by the solvable $T\bar T+\Lambda_2$ deformation with explicit Hamiltonian (up to matter-dependent fine structure among the cosmic horizon microstates that is not visible in the gravity description).  It would be interesting to connect to other proposals under development for finite quantum systems with de Sitter features such as \cite{Anninos:2022ujl, Anninos:2020geh, Rahman:2024vyg, Narovlansky:2023lfz}, including the method of trading fluxes in explicit compactifications for branes which reveal some of their microscopic  content \cite{Silverstein:2003jp}, applying that to modern examples such as \cite{Dong:2010pm, DeLuca:2021pej, Demirtas:2021ote}.  

In general, the non-gravitational timelike boundary included in this work simplifies holographic duality. 
It is of course essential to upgrade to four external spacetime dimensions (with uplift examples starting from AdS/CFT available in \cite{DeLuca:2021pej}).  There, at least naively we lose one of our essential ingredients: the 2d $T\bar T$ prescription for regulating (cutting off) the theory.  
The obstruction lies in the graviton sector of the 4d gravity theory, which is absent in 3d.  This 4d graviton sector is analogous to the bulk matter in 3d.   
Perhaps it is possible to define a solvable quadratic operator even in 4d which only provides a Dirichlet condition for the non-graviton gravitational degrees of freedom. 
As in the present work, one would obtain this regulator via a very short initial segment of the trajectory, after which the full quadratic operator \cite{Hartman:2018tkw, Taylor:2018xcy} would enter (analogously to $(\mathcal{OO})_r$ in the present work).
Also in 4d, the subtleties with Dirichlet boundary conditions may require switching to other options such as conformal boundary conditions as in \cite{Anderson:2006lqb, Witten:2018lgb, Anninos:2023epi, Anninos:2024wpy, Liu:2024ymn}.\footnote{We note, however, that the Dirichlet approach to this would not require the well-definedness of Dirichlet conditions for an arbitrary boundary geometry.}     

Let us finish with one final comment.  In addition to simplifying holography, manifolds with boundary are as generic (perhaps more so) than the case without. However, such holes in space might be expected to have been diluted by inflation, similarly to monopoles and other exotica.  In any case, 
independently of theoretical motivations, the observed positive cosmological constant and other data is presumably compatible with the existence of such boundaries at some level, provided they are sufficiently small and of sufficiently low density.  It might be interesting to develop signatures -- such as novel types of mirrored images that are not caused by gravitational lensing -- and use observations to place bounds on the presence of non-gravitational timelike boundaries within our observable horizon (in the spirit of e.g. \cite{Bond:1999tf, Cornish:2003db}).

\noindent{\bf Acknowledgements}

We would like to thank D. Freedman and J. Aguilera-Damia for early collaboration on GR + EFT in bounded regions and V. Shyam,  R. Soni, B. Banihashemi and E. Shaghoulian for many discussions on aspects of the duality with matter.   We are grateful to O. Aharony, A. Ahmadain,   D. Anninos,  F. Denef, V. Gorbenko, M. Guica, R. Khan, H. Maxfield, G. Pennington, S. Shenker, D. Stanford, L. Susskind, H. Verlinde, and A. Wall for very useful discussions.  
This research is supported by a Simons Investigator award and National Science Foundation grant PHY-2310429. GT is supported by
CONICET (PIP grant 11220200101008CO), ANPCyT (PICT 2018-2517), CNEA, and Instituto Balseiro.

\appendix

\bibliographystyle{JHEP}
\bibliography{refs.bib}
\end{document}